\documentclass[prd,aps,twocolumn,floatfix,superscriptaddress,preprintnumbers,nofootinbib]{revtex4-2}

\usepackage{svg}
\usepackage[normalem]{ulem} 
\usepackage{graphicx,color}
\usepackage{bm}
\usepackage{amsmath}
\usepackage{amssymb}    
\usepackage{pifont}
\usepackage{cancel}

\usepackage{standalone}
\usepackage{slashed}
\usepackage{multirow}
\usepackage{hhline}
\usepackage{colortbl}
\usepackage{accents} 

\newcommand{\nn}{\nonumber}

\newcommand{\Tr}{\mathrm{Tr}}

\renewcommand{\(}{\left(}
\renewcommand{\)}{\right)}
\renewcommand{\[}{\left[}

\renewcommand{\vec}[1]{\bm{#1}}

\begin{document} 

\title{Factorization theorem for quasi-TMD distributions with kinematic power corrections}

\date{\today}
\newcommand*{\UCM}{Departamento de F\'isica Te\'orica \& IPARCOS, Universidad Complutense de Madrid, Plaza de Ciencias 1, E-28040 Madrid, Spain}\affiliation{\UCM}

\author{Alejandro Bris Cuerpo}\affiliation{\UCM}
\author{Arturo Arroyo-Castro}\affiliation{\UCM}
\author{Alexey Vladimirov}\affiliation{\UCM}

\begin{abstract}
We derive the factorization theorem for the quasi–transverse-momentum-dependent (quasi-TMD) correlator, including kinematic power corrections to all orders. The resulting expression involves only twist-two TMD distributions and is frame invariant. As in the leading-power approximation, the reduced soft factor factorizes multiplicatively; however, in contrast, the TMD evolution factor is not multiplicative but enters as a convolution with the nonperturbative TMD distribution. We present a numerical estimate of the difference between the derived expression and the leading-power approximation, finding it to be of the order of tens of percent for current lattice simulations. We also demonstrate that the inclusion of these corrections improves the agreement between phenomenological extractions of the Collins–Soper kernel and lattice results.
\end{abstract} 

\preprint{IPARCOS-UCM-26-011} 

\allowdisplaybreaks

\maketitle 
\section{Introduction}
\label{sec:introduction}

Lattice simulations of nonlocal QCD operators offer a unique opportunity to explore parton dynamics in regimes inaccessible to collider experiments~\cite{Ji:2013dva, Ma:2014jla, Constantinou:2022yye}. Corresponding approaches have been formulated for a variety of parton distributions, including collinear parton distributions~\cite{Radyushkin:2017cyf, Izubuchi:2018srq, Ji:2020ect}, transverse-momentum-dependent (TMD) parton distributions~\cite{Ji:2018hvs, Ji:2019ewn, Ebert:2019okf, Vladimirov:2020ofp, Ebert:2022fmh}, generalized parton distributions~\cite{Constantinou:2020hdm, Bhattacharya:2022aob}, and others. All these approaches share a common feature: they are based on factorization theorems in the hadron momentum $P_z$, and thus the precision of the theoretical description depends crucially on the value of $P_z$. At present, lattice simulations can reach momenta of about $P_z \simeq 3$~GeV~\cite{Bali:2016lva}, which is only moderately larger than the nonperturbative QCD scale. Therefore, one may expect sizable power corrections. In this work, we investigate the magnitude of power corrections for quasi-TMD distributions (quasi-TMDs) and explore their impact on the extraction of TMD distributions.

Quasi-TMD correlators are lattice observables used to determine transverse-momentum-dependent parton distribution functions (TMDPDFs) and the Collins–Soper (CS) kernel, which are central elements of nucleon tomography in momentum space~\cite{Angeles-Martinez:2015sea, Boussarie:2023izj}. They are defined as matrix elements of spatially separated quark and antiquark fields connected by a staple-shaped Wilson line of finite length~\cite{Ebert:2022fmh}. In the limit $P_z \to \infty$, these correlators can be expressed in terms of the standard TMDPDFs and the so-called reduced soft factor~\cite{Ji:2018hvs, Ji:2019ewn, Ebert:2019okf, Vladimirov:2020ofp, Rodini:2022wic} through a QCD factorization theorem. This relation has been exploited by several groups to determine the CS kernel~\cite{LatticeParton:2020uhz, Schlemmer:2021aij, Shanahan:2021tst, Li:2021wvl, LatticePartonLPC:2022eev, Shu:2023cot, Avkhadiev:2023poz, Avkhadiev:2024mgd, Bollweg:2024zet, Alexandrou:2025xci, Tan:2025ofx} as well as the TMDPDFs themselves~\cite{LatticePartonCollaborationLPC:2022myp, Bollweg:2025iol}. These simulations have demonstrated that this approach complements traditional phenomenological extractions of TMD distributions~\cite{Bacchetta:2017gcc, Scimemi:2017etj, Bertone:2019nxa, Scimemi:2019cmh, Vladimirov:2019bfa, Bury:2020vhj, Moos:2023yfa, Bacchetta:2022awv, Boglione:2022nzq, Barry:2023qqh, Boglione:2023duo, Bacchetta:2024qre, Barry:2025glq}, providing access to the large-$b$ region~\cite{Avkhadiev:2025wps}.

In essence, the factorization theorem for quasi-TMDs corresponds to the standard TMD factorization theorem, with the hard scale provided by the hadron momentum $P_z$ (or, more precisely, by the momentum of the leading parton, $xP_z$). Owing to its intrinsic multidimensional structure, the TMD factorization theorem exhibits a nontrivial pattern of power corrections~\cite{Balitsky:2020jzt, Vladimirov:2021hdn, Vladimirov:2023aot}. It is convenient to classify these corrections into the following four categories:
\begin{itemize}
\item \textit{Kinematic power corrections (KPCs):} These corrections arise due to the nonzero own transverse momentum of partons and appear as derivatives of TMD matrix elements. They are generically suppressed by a factor $\sim |\vec k_T|/P_z$, where $|\vec k_T|$ denotes the parton transverse momentum.

\item \textit{Genuine higher-twist corrections:} These corrections describe multiparton dynamics that emerge at subleading power and are parametrized in terms of genuine higher-twist distributions. They are generically suppressed by $\sim \Lambda/P_z$, where $\Lambda$ is the nonperturbative QCD scale. Although higher-twist TMD distributions are largely unknown in practice, they are expected to be orders of magnitude smaller than twist-two distributions, in analogy with the collinear case~\cite{Vladimirov:2025qrh}.

\item \textit{$q_T$-corrections:} These corrections arise in the limit of small transverse separation between the factorized currents and are characterized by the scale $1/(bP_z)$~\cite{Arroyo-Castro:2025slx}. This type of correction governs the transition region between TMD and collinear factorization. However, these corrections can be controllably neglected in lattice computations, where the parameter $b$ is directly specified.

\item \textit{Target-mass corrections:} These corrections arise due to the finite hadron mass and scale as $M/P_z$. One therefore expects them to be numerically significant for current lattice simulations (see, for instance,~\cite{Braun:2018brg}).
\end{itemize}
At present, the quasi-TMD correlator is known up to next-to-leading power (NLP)~\cite{Rodini:2022wic, Rodini:2022wki}, and the higher-order power corrections remain unexplored.

In this work, we analyze kinematic power corrections (KPCs) in the quasi-TMD factorization theorem. We argue that these corrections are likely to be the most important among all types of power corrections, both theoretically and numerically. Their numerical dominance cannot be firmly established without a complete analysis of all contributions. However, we expect genuine higher-twist corrections to be relatively small, since they are proportional to multiparton correlation functions (whereas KPCs are expressed in terms of standard TMD distributions), while $q_T$-corrections are controlled by the lattice setup and are negligible in the large-$b$ region. One should also be aware that target-mass corrections may be significant; however, no systematic method for computing them is currently available, and they are typically accompanied by an additional factor of $x^2$. Therefore, under present conditions, KPCs represent the most relevant class of power corrections to investigate.

The KPCs play an important role in the TMD factorization theorem, as they restore frame invariance (or reparameterization invariance~\cite{Manohar:2002fd}) and other symmetries that are broken at leading power (LP) (such as the charge conservation for the Drell-Yan hadron tensor \cite{Balitsky:2020jzt, Piloneta:2024aac}). In this sense, KPCs are not independent power corrections. Rather, they are descendants of the LP term and can be partially deduced from it. For reviews of KPCs in TMD and collinear factorization, we refer to refs.~\cite{Vladimirov:2023aot, Piloneta:2024aac, Braun:2011dg, Braun:2011zr}. Another notable property of KPCs is that they can be computed and resummed exactly to all orders in the power expansion~\cite{Vladimirov:2023aot}, leading to an improved formulation of the TMD factorization theorem, referred to as the TMD-with-KPC framework~\cite{Piloneta:2025jjb}. This approach is as fundamental as the traditional TMD factorization, sharing all its key ingredients -- such as evolution and coefficient functions -- while additionally restoring symmetries broken at LP approximation. In the case of the Drell–Yan process and semi-inclusive deep-inelastic scattering (SIDIS), KPCs have been shown to be numerically significant for data with $Q \lesssim 10$~GeV~\cite{Piloneta:2024aac, Piloneta:2025jjb}.

Following refs.~\cite{Vladimirov:2023aot, Piloneta:2024aac}, we derive the TMD-with-KPC expressions for quasi-TMD matrix elements, restricting ourselves to those with nonvanishing leading-power (LP) contributions, i.e., those that do not involve mixing with twist-three TMD distributions at LP. The derivation is presented in Sec.~\ref{sec:derivation}. In Sec.~\ref{sec:numImpact}, we estimate the impact of KPCs on the determination of the CS kernel and TMDPDFs from current lattice setups. The conclusions are presented in Sec.~\ref{sec:conclusions}. 

\section{Expression for quasi-TMD in TMD-with-KPC formalism}
\label{sec:derivation}

In this section, we derive the factorization theorem for the quasi-TMD matrix element, including kinematic power corrections (KPCs) that follows the leading-power (LP) term to all powers. The derivation closely follows analogous treatments of factorization theorems for scattering cross sections~\cite{Vladimirov:2023aot, Piloneta:2024aac, Piloneta:2025jjb}. For this reason, we do not present the technical aspects of the derivation in full detail, but instead focus on their consequences.

\textbf{Definitions.} 
The quasi-TMD matrix element is defined as following \cite{Ebert:2022fmh, Vladimirov:2020ofp, Ji:2018hvs}
\begin{eqnarray}
&&\widetilde{\Omega}^{[\Gamma]}_{q/h}(y;\mu)= Z^{-1}\times
\\\nn && \langle P,s|\bar q(y)[y,b+Lv][b+Lv,Lv][Lv,0]\frac{\Gamma}{2}q(0)|P,s\rangle
\end{eqnarray}
where $b^\mu=y_T^\mu=y^\mu-v^\mu (yv)/v^2$, $\langle P,s|$ is a hadron state with momentum $P$ and spin $s$, $q$ are quark fields, $\Gamma$ is a Dirac matrix, and $[a_1,a_2]$ denotes the straight gauge link between points $a_1$ and $a_2$. The renormalization scale $\mu$ enters via the renormalization constant, which in the continuum is defined as
\begin{equation}\label{def:Z}
Z^{-1}=Z_W^{-1}(y,L,\mu)Z_J^{-2}(\mu),
\end{equation}
where $Z_W$ is the renormalization factor for the gauge links, and $Z_J$ is the renormalization factor for quark fields (in the axial gauge).

Both vectors $y$ and $v$ are purely spatial, such that this matrix element is suitable for lattice simulations. The momentum of hadron defines standard light-cone directions
\begin{eqnarray}
P^\mu=P^+ \bar n^\mu+\frac{M^2}{2P^+}n^\mu,
\end{eqnarray}
where $n^2=\bar n^2=0$, $(n\bar n)=1$, and $P^2=M^2$ is the mass of hadron, in the following considered to be negligible. For simplicity, we orient the operator such that the vector $v$ belongs to the plane $(n,\bar n)$\footnote{
If the vector $v^\mu$ has transverse components (i.e., $(v\cdot b)\neq 0$), the derivation of the factorization theorem is analogous, although it requires the assumption $|b|\gg |v\cdot b|$. The final form is the same as in (\ref{tree-order}), with the argument of the reduced soft factor $\Psi$ being $(b - v_T y^-/v^-)$. If the condition $|b|\gg |v\cdot b|$ is violated, the factorization theorem gets harmful power corrections; see the discussion in Ref.~\cite{Ebert:2022fmh}. Here we consider only the case (\ref{def:v}), as it is the simplest and the most commonly used in practice.
}, and can be written as
\begin{eqnarray}\label{def:v}
v^\mu=\frac{n^\mu-\bar n^\mu}{\sqrt{2}},\qquad v^2=-1,\qquad (v\cdot P)=P_v>0.
\end{eqnarray}
In this case, the vector $y$ can be decomposed as
\begin{eqnarray}
y^\mu=\ell v^\mu+b^\mu,
\end{eqnarray}
such that $(b\cdot v)=(b\cdot P)=0$ and $\ell=(yv)/v^2$.

The TMD factorization theorem can be applied in the following limit:
\begin{eqnarray}
|L|\gg |y|,\qquad P_+\gg \Lambda,\qquad |b|\gg \Lambda^{-1},
\end{eqnarray}
where $\Lambda$ is a generic low-energy scale of QCD. These limits have distinct implications with respect to the factorization theorem and power corrections. Namely,
\begin{itemize}
\item The limit $|L|\gg |y|$ is required to suppress the interaction between the ``far'' end of the staple gauge link and the quark currents. This limit is essential for the entire factorization approach, because it allows one to separate the soft gluon radiation from the hard radiation. Most probably, power corrections with respect to $L$ violate the factorization theorem. For that reason, this limit is taken prior to any other limit.
\item The limit $P_+\gg \Lambda$ is required to suppress genuine power corrections and target-mass corrections.
\item The limit $|b|\gg \Lambda^{-1}$ corresponds to the small-transverse-momentum limit in the ordinary factorization. It is required to suppress $q_T$-corrections.
\end{itemize}
Note that none of these limits is related to KPCs. The KPCs are related to the transverse momenta of partons, $k_T$, which are not controlled by the external kinematics, nor have a fixed value. It is often assumed that $k_T\sim \Lambda$ GeV, and thus KPCs are suppressed to the same extent as genuine and target-mass corrections. However, this is not entirely correct, since the value of $k_T$ is integrated over an infinite range. Due to the shape of nonperturbative TMD distributions, $k_T$ can take certain effective values, and practical studies~\cite{delRio:2024vvq, Piloneta:2024aac, Piloneta:2025jjb} show that this effective value strongly depends on $x$ and is of the order of $k_T\sim 2$--$3$ GeV for moderate values of $x$.

Conventionally, one also defines the quasi-TMD correlator in the momentum-fraction space as
\begin{eqnarray}\label{omega-x}
\widetilde \Omega^{[\Gamma]}_{q/h}(x,b;\mu)=\int_{-\infty}^\infty \frac{d \ell}{2\pi}e^{-ix\ell P_v}\widetilde{\Omega}^{[\Gamma]}_{q/h}(\ell v+b;\mu).
\end{eqnarray}

\textbf{Factorized expression at the tree-order.} 
The derivation of the TMD factorization theorem at higher powers can be most conveniently performed within the background field method~\cite{Vladimirov:2021hdn}. Its application to quasi-TMD matrix elements is presented in detail in Ref.~\cite{Rodini:2022wic}. The key element of the higher-power factorization is the decomposition of TMD correlators in terms of distributions with specific TMD-twists. The TMD-twist is defined such that TMD distributions with distinct TMD-twists do not mix with each other during the evolution procedure, and thus are completely independent. As a result, a series of power corrections with operators of a given TMD-twist can be considered independently of other series. It obeys all the symmetries and properties of the original operator.

The KPCs then correspond to the series of terms with TMD-twist-two only. This limitation allows one to identify the KPC contribution at all powers, because TMD-twist-two operators are those that contain only two good components of the quark field and no additional gluon fields (except for the gluon fields that compose the gauge links). Therefore, to extract the KPCs, it is sufficient to compute the factorization theorem with vanishing background gluon fields, and re-express the bad components of the remaining fields in terms of the good ones using the QCD equations of motion.

In the absence of the background gluon fields, the tree-order factorization reduces to algebraic manipulations with indices, and the resulting expression reads 
\begin{eqnarray}
\widetilde{\Omega}^{[\Gamma]}_{q/h}(y)&=& \widetilde{\Phi}^{[\Gamma]}_{q/h}(y) \Psi(y)
\\\nn &&
+\text{(contributions with gluon fields)},
\end{eqnarray}
where $\widetilde{\Phi}^{[\Gamma]}_{q/h}(y)$ is the TMD correlator
\begin{eqnarray}\label{Phi-gen}
\widetilde{\Phi}^{[\Gamma]}_{q/h}(y)=\langle P,s|\bar q(y)\frac{\Gamma}{2}q(0)|P,s\rangle,
\end{eqnarray}
and $\Psi(y)$ is the ``reduced soft-factor''
\begin{eqnarray}
\Psi(y)=\frac{\Tr}{N_c}\langle 0|H(y) H^\dagger(0)|0\rangle,
\end{eqnarray}
with $H(y)=[y,b+Lv][b+Lv,\infty_T+Lv]$. Both expressions are presented in the light-cone gauge, and in a general gauge should be supplemented by appropriate light-cone gauge links.

The next step is to extract the leading TMD-twist component of the correlator (\ref{Phi-gen}). For the TMD correlator with a general separation between fields it has been done in ref.~\cite{Vladimirov:2023aot}. The result can be conveniently written in the form of integral in the momentum space
\begin{widetext}
\begin{eqnarray}\label{Psi:4}
\widetilde{\Phi}^{[1]}_{q/h}(y)\Big|_{\text{tw2}}&=&\widetilde{\Phi}^{[\gamma^5]}_{q/h}(y)\Big|_{\text{tw2}}=0,
\\
\widetilde{\Phi}^{[\gamma^\mu]}_{q/h}(y)\Big|_{\text{tw2}}&=&2P^+
\int d^4 k \,\delta(k^2)e^{i(ky)}\int d\xi\,\delta(k^+-\xi P^+)k^\mu \Phi_{11}^{[\gamma^+]}(\xi,k_T),
\\
\widetilde{\Phi}^{[\gamma^\mu\gamma^5]}_{q/h}(y)\Big|_{\text{tw2}}&=&2P^+
\int d^4 k \,\delta(k^2)e^{i(ky)}\int d\xi\, \delta(k^+-\xi P^+) k^\mu \Phi_{11}^{[\gamma^+\gamma^5]}(\xi,k_T),
\\\label{Psi:sigma}
\widetilde{\Phi}^{[i\sigma^{\mu\nu}\gamma^5]}_{q/h}(y)\Big|_{\text{tw2}}&=&2P^+
\int d^4 k \,\delta(k^2)e^{i(ky)}\int d\xi\, \delta(k^+-\xi P^+) 
\\\nn && 
\times\Big(g^{\mu\alpha}_T k^\nu-g^{\nu\alpha}_T k^\mu+\frac{k^\mu n^\nu-n^\mu k^\nu}{k^+}k_T^\alpha\Big)
\Phi_{11}^{[i\sigma^{\alpha +}\gamma^5]}(\xi,k_T),
\end{eqnarray}
\end{widetext}
where $\Phi_{11}$ are ordinary TMD distributions defined in the momentum space
\begin{eqnarray}
&&\Phi_{11}^{[\Gamma]}(\xi,k_T)=
\int \frac{d^2b}{(2\pi)^2}e^{-i(bk)_T}\widetilde{\Phi}^{[\Gamma]}_{11}(\xi,b)
\\\nn &&\qquad=\int \frac{dy^- d^2b}{(2\pi)^3}e^{-i\xi y^-P_+-i(bk)_T}\widetilde{\Phi}^{[\Gamma]}_{11}(y^-n+b).
\end{eqnarray}
Note, that the values of $\xi$ are restricted by the causality as $\xi\in[0,1]$.

\begin{table*}[t]
\centering
\renewcommand{\arraystretch}{2.1}
\begin{tabular}{|c|c|c|c|}
    \hline
    $\Gamma$ & $K^{[\Gamma]}$ & $K_\text{LP}^{[\Gamma]}$ & $\Gamma_+$ \\[2pt]
    \hline
    $1,\gamma^5$ & $0$ & $0$ & $\varnothing$ \\[2pt]
    \hline
    $\gamma^\mu$ & $k^\mu$ & $\sqrt{2}xP_v\bar n^\mu$ & $\gamma^+$ \\[2pt]
    \hline
    $\gamma^\mu\gamma^5$ & $k^\mu$ & $\sqrt{2}xP_v\bar n^\mu$ & $\gamma^+\gamma^5$ \\[2pt]
    \hline
    ~$i\sigma^{\mu\nu}\gamma^5$~ & ~$g^{\mu\alpha}_Tk^\nu-g^{\nu\alpha}_Tk^\mu+\dfrac{k^\mu n^\nu-n^\mu k^\nu}{k^+}k_T^\alpha$~ & ~$\sqrt{2}xP_v(g^{\mu\alpha}_T\bar n^\nu-g^{\nu\alpha}_T\bar n^\mu)$~ & ~$i\sigma^{\alpha+}\gamma^5$~ \\[2pt]
    \hline
\end{tabular}
\caption{\label{tab:K} The table of correspondence between elements of the Dirac basis $\Gamma$, and elements in the factorized expressions.}
\end{table*}

The leading-TMD-twist component of $\Psi$ can be derived analogously. In this case, the equation of motion is
\begin{eqnarray}\label{EOM:H}
(v\cdot D)H=\mathcal{O}\(\frac{1}{L}\).
\end{eqnarray}
In the absence of background gluon fields, it leads to
\begin{eqnarray}
\Psi_{\text{lead.tw.}}(y)&=&\Psi(b).
\end{eqnarray}
Therefore, the expression for TMD-with-KPC factorization \textit{at tree-order} is
\begin{eqnarray}\label{tree-order}
\widetilde{\Omega}^{[\Gamma]}_{q/h}(y)&=& 2P_+ \Psi(b)\int d\xi\int d^4k \,
\\\nn &&\times
\delta(k^+-\xi P^+) \delta(k^2)
e^{i(ky)}K^{[\Gamma]}\Phi_{11}^{[\Gamma_+]}(\xi,k_T),
\end{eqnarray}
where the expression for $K^{[\Gamma]}$ and corresponding $\Gamma_+$ are given in the table \ref{tab:K}.

\textbf{Factorized expression beyond the tree-order.} 
Beyond the tree order, the factorized expression receives the coefficient function and the dependence on the factorization scales. At one loop, it is produced by a single diagram~\cite{delRio:2023pse}. Performing the computation of this diagram in general kinematics, i.e., without applying the power expansion, we obtain the following result
\begin{eqnarray}\label{1-loop}
&&H^\dagger(0)q(0)\Big|_{\text{1-loop}}=
2a_s C_F \Gamma(-\epsilon)\Gamma(2\epsilon)\frac{1-\epsilon}{1-2\epsilon}
\\\nn &&\qquad\times
\(-\frac{v^2}{4}\)^\epsilon 
\frac{1}{[(v\cdot \partial)-s0]^{2\epsilon}} H^\dagger(0)q(0),
\end{eqnarray}
where $s=\text{sign}(L)$ and $a_s=g^2/(4\pi)^{d/2}$ is the QCD coupling constant. This expression coincides with the NLP computation~\cite{Rodini:2022wic}. Expression~(\ref{1-loop}) proves that the coefficient function is the same for KPCs at all powers, which is a fundamental property of the KPCs guaranteed by Lorentz invariance.

The important observation is that the derivative $(v\cdot \partial)$ does not act on $H$, because of the equation of motion (\ref{EOM:H}). This derivative acts solely on the quark field and produces the momentum of the parton. It also defines the Lorentz-invariant argument of the hard coefficient function as $4(v\cdot k)^2$, where $k$ is the parton momentum defined within the Fourier transform (\ref{tree-order}). Note that, at LP, this argument is $(2x(v\cdot P))^2$, which corresponds to the $k_+$-only part of the Lorentz-invariant argument. This is the only modification of the coefficient function within the KPC formalism. This modification makes the factorized expression Lorentz invariant and has important consequences.

The singularities of the coefficient function are renormalized by the factor $Z^{-1}$ (\ref{def:Z}) and the renormalization factors of $\Phi$ and $\Psi$, in a complete analogy of LP computation. The only modification brought by KPC is the condition for the rapidity renormalization scales $\zeta$ and $\bar \zeta$. This condition is modified due to the change in the argument of the coefficient function and becomes
\begin{eqnarray}\label{zeta-bar-zeta}
\zeta \bar \zeta=\mu^2 (2v\cdot k)^2,
\end{eqnarray}
where $\mu$ is the factorization scale. The evolution equations with respect to the rapidity scale are
\begin{eqnarray}\label{ev:Phi}
\zeta \frac{\partial}{\partial \zeta}\widetilde\Phi_{11}^{[\Gamma]}(x,b;\mu,\zeta)&=&-\mathcal{D}(b,\mu)\widetilde\Phi_{11}^{[\Gamma]}(x,b;\mu,\zeta),
\\\label{ev:Psi}
\zeta \frac{\partial}{\partial \zeta}\Psi(b;\mu,\zeta)&=&-\mathcal{D}(b,\mu)\Psi(b;\mu,\zeta),
\end{eqnarray}
where $\mathcal{D}$ is the Collins-Soper kernel (also often defined as $K=-2\mathcal{D}$).

There is one last important comment to be made. The TMD factorization beyond LP possesses the phenomenon of special rapidity singularities~\cite{Rodini:2022wki}. They cancel between various terms of the factorization theorem at a given power, but leave a trace in the form of logarithms $\ln(\zeta/\bar \zeta)$ accompanied by derivatives of the Collins–Soper kernel. These terms add up together with derivatives in KPCs and form boost-invariant derivatives. This immediately leads to a severe complication, namely, the series of KPCs cannot be summed into simple expressions (\ref{Psi:4} - \ref{Psi:sigma}) unless $\zeta=\bar \zeta$ (such that boost-invariant logarithms vanish). The distributions however, still obey the evolution equations (\ref{ev:Phi}, \ref{ev:Psi}), and thus one can change the rapidity scale to any convenient point after the summation of KPCs, which lead to rather involved expressions. Therefore, in what follows, we consider the case $\zeta=\bar \zeta$ for simplicity. 

\textbf{Final expression for the quasi-TMD correlator.} 
Summarizing these observations we write the final form for the quasi-TMD correlator:
\begin{widetext}
\begin{eqnarray}
\widetilde{\Omega}^{[\Gamma]}_{q/h}(y;\mu)&=& 2P_+ \Psi(b;\mu,\bar \zeta)\int d\xi\int d^4k\, \delta(k^+-\xi P^+) \delta(k^2)e^{i(ky)}C\(\frac{2k_v}{\mu}\)
K^{[\Gamma]}\Phi_{11}^{[\Gamma_+]}(\xi,k_T;\mu,\zeta),
\end{eqnarray}
where $k_v=(v\cdot k)$, $\zeta=\bar \zeta$ and their product is given by equation (\ref{zeta-bar-zeta}). The coefficient function is the LP coefficient function at all perturbative orders (currently known up to next-to-next-to-leading order (NNLO) \cite{delRio:2023pse, Ji:2023pba}). 

In the momentum-fraction representation (\ref{omega-x}) it reads 
\begin{eqnarray}
\widetilde{\Omega}^{[\Gamma]}_{q/h}(x,b;\mu)&=&2P_+ \Psi(b;\mu,\bar \zeta)
\\\nn &&\times 
\int d\xi\int d^4k \,\delta(k^+-\xi P^+) \delta(k_v-xP_v) \delta(k^2)e^{-i(\vec k\vec b)}
C\(\frac{2k_v}{\mu}\)
K^{[\Gamma]}\Phi_{11}^{[\Gamma_+]}(\xi,k_T;\mu,\zeta),
\end{eqnarray}
where the scalar product of transverse components is written with the Euclidean norm $(\vec k\vec b)=-(kb)$. Here and in the following, we follow the standard practice to indicate by the bold font the transverse components of the vectors with the Euclidian norm. Integrating over $k_{0,v}$ and $\xi$ the expression simplifies
\begin{eqnarray}\label{main}
\widetilde{\Omega}^{[\Gamma]}_{q/h}(x,b;\mu)&=& \Psi(b;\mu,\bar \zeta)C\(\frac{2xP_v}{\mu}\)\int d^2k_T \,e^{-i(\vec k\vec b)}
\frac{K^{[\Gamma]}}{\sqrt{x^2P_v^2+\vec k_T^2}}\Phi_{11}^{[\Gamma_+]}(\xi,k_T;\mu,\zeta),
\end{eqnarray}
where
\begin{eqnarray}\label{KPC-kinematic}
\xi=\frac{x}{2}\(1+\sqrt{1+\frac{\vec k_T^2}{x^2P_v^2}}\),\qquad k_v=xP_v,\qquad k_0=\sqrt{x^2P_v^2+\vec k_T^2}.
\end{eqnarray}
\end{widetext}
The value of $\xi$ is restricted to $\xi\in[x,1]$, which in turn, restricts the integration range of $k_T$ as $|\vec k_T|<2P_v\sqrt{1-x}$. Note, that the coefficient function is moved out of the $k_T$-integral, and the values of $\zeta$'s are given by
\begin{eqnarray}\label{zeta=zeta}
\zeta=\bar \zeta=2x\mu P_v.  
\end{eqnarray}

In the limit $(xP_v)^2\gg \vec k^2_T$ the expression (\ref{main}) reproduces the standard factorization theorem for quasi-TMD correlator. This limit must be taken prior the integration, and thus $\xi$ turns to $x$, and the factor $K^{[\Gamma]}$ becomes dominated by $k_v=xP_v$. Then the Fourier transform is trivially taken and one obtains
\begin{eqnarray}\label{LP-final}
\lim_{\text{LP}}\widetilde{\Omega}^{[\Gamma]}_{q/h}(x,b;\mu)
&=&
\Psi(b;\mu,\bar \zeta)
\\\nn && \times C\(\frac{2xP_v}{\mu}\)
\frac{K_{\text{LP}}^{[\Gamma]}}{xP_v}\widetilde{\Phi}_{11}^{[\Gamma_+]}(x,b;\mu,\zeta),
\end{eqnarray}
where $K_{\text{LP}}^{[\Gamma]}$ is the LP limit of $K^{[\Gamma]}$ presented in the table \ref{tab:K}. This is the well-known expression for LP factorization of quasi-TMD matrix element. Note, that in the position space the requirement (\ref{zeta=zeta}) is not important, because the evolution is multiplicative.

The expression depends on the particular value of $\Gamma$, via the coefficient $K^{[\Gamma]}$. For example, for the unpolarized hadron after the integration over the angle of transverse momentum one finds
\begin{eqnarray}\label{unpol1}
&&\widetilde{\Omega}^{[\gamma^0]}_{q/h}(x,b;\mu)= 2\pi\Psi(b;\mu,\bar \zeta)C\(\frac{2xP_v}{\mu}\)
\\\nn &&\qquad \times 
\int_0^{2P_v\sqrt{1-x}} d|k| |k| J_0(|k|b)
f_1(\xi,|k|;\mu,\zeta),
\\\label{unpol2}
&&\widetilde{\Omega}^{[\gamma^v]}_{q/h}(x,b;\mu)= 2\pi\Psi(b;\mu,\bar \zeta)C\(\frac{2xP_v}{\mu}\)
\\\nn &&\qquad \times 
\int_0^{2P_v\sqrt{1-x}} d|k| |k| J_0(|k|b)
\frac{2\xi-x}{x}f_1(\xi,|k|;\mu,\zeta),
\end{eqnarray}
where $f_1$ is the unpolarized TMD distribution and $J_0$ is the zeroth order Bessel function of the first kind. The explicit expressions for other components, including spin-dependent TMD distributions, have similar form.

In the practice, one often consider the reduced quasi-TMD correlator, that is obtained by removing the reduced soft factor at the scale $\mu$, i.e. by dividing (\ref{main}) by $\Psi(b;\mu,\mu^2)$. The corresponding expression can be obtained using the evolution equation (\ref{ev:Psi})
\begin{eqnarray}\label{main2}
&&\frac{\widetilde{\Omega}^{[\Gamma]}_{q/h}(x,b;\mu)}{
\Psi(b;\mu,\mu^2)}
= C\(\frac{2xP_v}{\mu}\)\(\frac{2xP_v}{\mu}\)^{-\mathcal{D}(b,\mu)}
\\\nn &&\qquad \times 
\int d^2k_T \,e^{-i(\vec k\vec b)}
\frac{K^{[\Gamma]}}{\sqrt{x^2P_v^2+\vec k_T^2}}\Phi_{11}^{[\Gamma_+]}(\xi,k_T;\mu,\zeta).
\end{eqnarray}
In this form, the factorization formula is very similar to the LP expression (compare e.g. with refs.~\cite{Ebert:2022fmh, Ji:2019ewn, Rodini:2022wic, Vladimirov:2020ofp}). The only difference is the value $\xi$ that depends on the parton's momentum. Such modification is welcome because the LP factorization theorem allows the parton to have any transverse momentum, including infinitely large, while the expression (\ref{main}) respects natural kinematical constraints.

The expression (\ref{main}) reveals the dependence of power corrections on $x$. Clearly, these corrections are enhanced at small and large $x$. At smaller $x$ they behave as $\mathcal{O}(\vec k_T^2/(xP_v)^2)$, which is a known behavior, see \cite{Ebert:2022fmh, Vladimirov:2020ofp}. However, the causality restriction $|\xi|<1$ also leads to corrections of order $\mathcal{O}(\vec k_T^2/(1-x)P_v^2)$, which were not discussed earlier. Let us mention that similar corrections (although from a different origin) arise in the factorization of quasi-PDF matrix elements \cite{Braun:2018brg, Liu:2023onm}.

\begin{figure*}[th]
\centering
\includegraphics[width=0.95\textwidth]{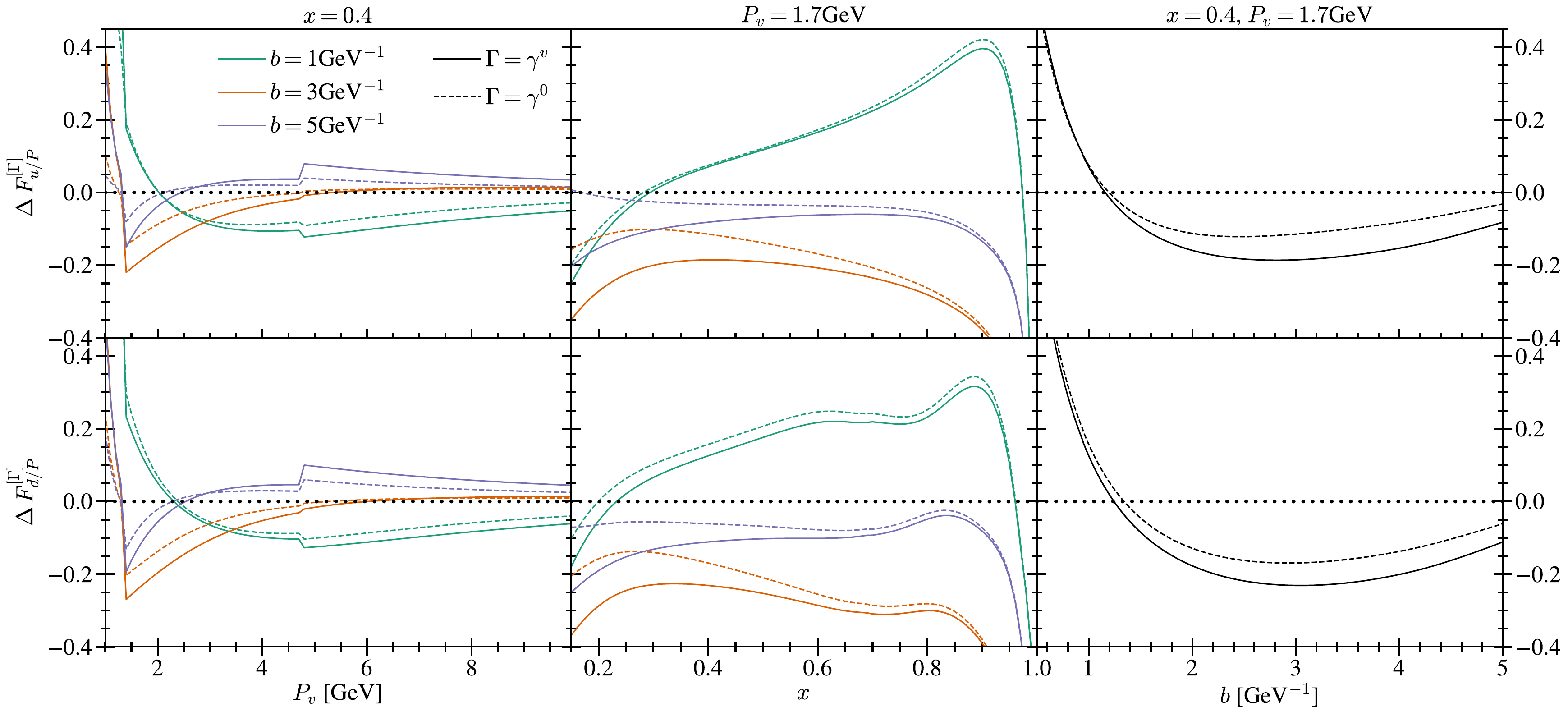}
\caption{Relative difference between the reduced unpolarized quasi-TMDPDF with KPCs and at LP: $\Delta\,F^{[\Gamma]}_{q/P}=(F^{[\Gamma],\,\rm{KPC}}_{q/P}-F^{[\Gamma],\,\rm{LP}}_{q/P})/F^{[\Gamma],\,\rm{LP}}_{q/P}$, as a function of proton's momentum (left), parton's momentum fraction $x$ (center) and transverse distance $b$ (right) for $u$ (up panels) and $d$ (down panels) quarks. The renormalization scale is chosen to be equal to the proton momentum $\mu=P_v$, and the remaining variables are fixed to those values indicated at the top of each column. The comparison is done using the central values of ART25 extraction \cite{Moos:2025sal}. The kinks in $P_v$ plots correspond to charm and bottom mass scales.}
\label{fig:RelDiff}
\end{figure*}

Evidently, the main modification brought by KPCs is the dependence of the parton’s momentum fraction $\xi$ on its transverse momentum $\vec k_T$. This modification prevents multiplicative factorization of the quasi-TMD in transverse position space, as occurs at LP. It also implies that the dependence on $\zeta$ is not multiplicative, and one cannot single it out by taking the ratio of quasi-TMD correlators. In the following section, we analyze this expression numerically.

\section{Numerical significance of KPCs}
\label{sec:numImpact}

The evaluation of the numerical significance of KPCs is not straightforward, since they are not presented as an additional or multiplicative factor, but instead modify the form of the integral convolution. Therefore, they essentially depend on the shape of the test function. In this section, we estimate the size of the correction using the ART25 TMD extraction~\cite{Moos:2025sal} as the test input. The quasi-TMD coefficient function is taken at NNLO~\cite{delRio:2023pse}.

To make a comparison clearer we consider the reduced quasi-TMD correlator defined as
\begin{eqnarray}
F^{[\Gamma]}_{q/h}(x,b;\mu)\equiv\frac{\widetilde{\Omega}^{[\Gamma]}_{q/h}(x,b;\mu)}{
\Psi(b;\mu,\mu^2)}\,.
\end{eqnarray}
On the one hand, this expression can be measured using lattice QCD (see, for instance, \cite{Bollweg:2025iol, LatticePartonCollaborationLPC:2022myp}). On the other hand, there is the factorization theorem derived in the previous section. The complete (i.e., Lorentz-invariant) form of the factorization theorem is given in (\ref{main}).
\begin{eqnarray}\label{F=KPC}
&&F^{[\Gamma],\text{KPC}}_{q/h}(x,b;\mu)=C\(\frac{2xP_v}{\mu}\)\(\frac{2xP_v}{\mu}\)^{-\mathcal{D}(b,\mu)}
\\\nn &&\quad \times 
\int d^2k_T \,e^{-i(\vec k\vec b)}
\frac{K^{[\Gamma]}}{\sqrt{x^2P_v^2+\vec k_T^2}}\Phi_{11}^{[\Gamma_+]}(\xi,k_T;\mu,\zeta),
\end{eqnarray}
while the LP approximation is
\begin{eqnarray}\label{F=LP}
&&F^{[\Gamma],\,\text{LP}}_{q/h}(x,b;\mu)=
C\(\frac{2xP_v}{\mu}\)\(\frac{2xP_v}{\mu}\)^{-\mathcal{D}(b,\mu)}
\\\nn &&\quad \times 
\frac{K^{[\Gamma]}_{\text{LP}}}{xP_v}\widetilde{\Phi}_{11}^{[\Gamma_+]}(x,b;\mu,\zeta)\,.
\end{eqnarray}
In both cases, $\zeta=2\mu x P_v$. In what follows, we consider only the cases $\Gamma=\gamma^0$ and $\Gamma=\gamma^v$ with a spinless hadron, which project to the unpolarized TMDPDF, see (\ref{unpol1}, \ref{unpol2}).

\begin{figure*}[t]
\centering
\includegraphics[width=0.95\textwidth]{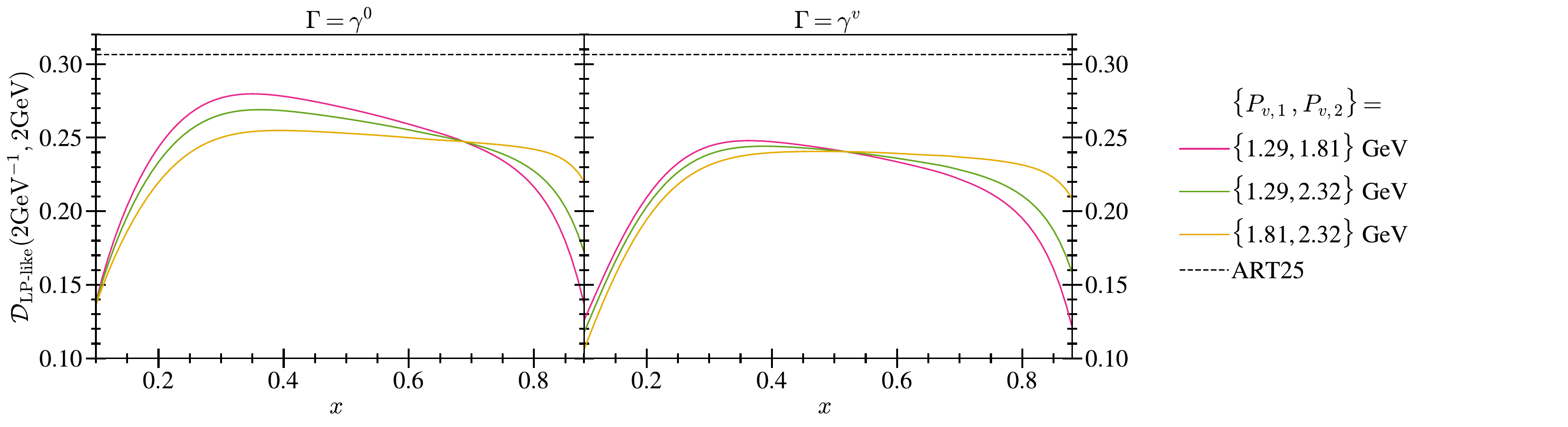}
\caption{CS kernel at $b=2$ GeV$^{-1}$ and $\mu=2$ GeV, obtained from eq.~(\ref{CS}) for different values of $x$ and $\Gamma$, computed using the central values of ART25. The solid lines correspond to various combinations of hadron momenta considered in lattice data extraction~\cite{Avkhadiev:2024mgd}. The ART25 input is shown as a dashed line.}
\label{fig:CSxdependence}
\end{figure*}

\begin{figure*}[t]
\centering
\includegraphics[width=0.95\textwidth]{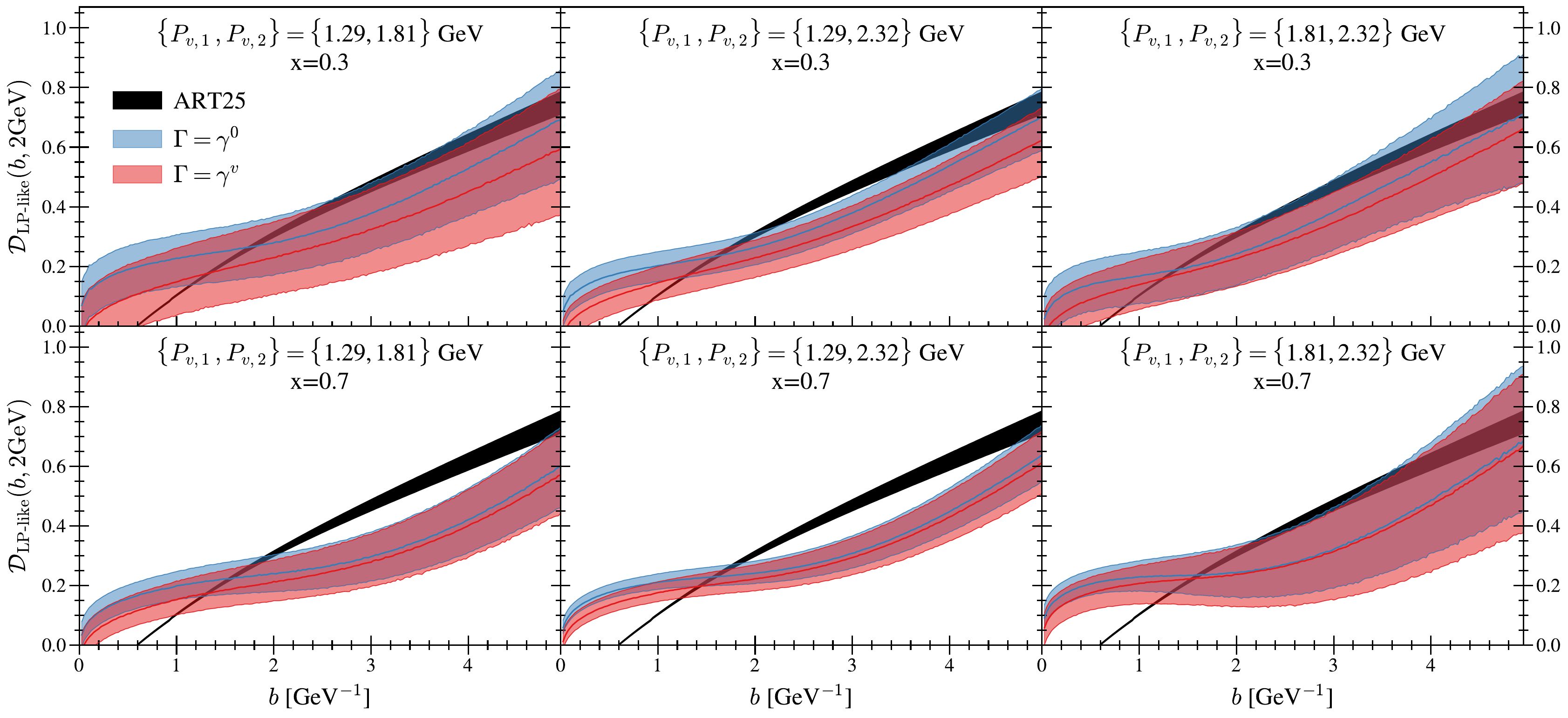}
\caption{CS kernel at $\mu=2$ GeV, obtained from eq.~(\ref{CS}) using the ART25 input. The employed combinations of hadron momenta are indicated at the top of each panel, together with the value of $x$. The black band represents the CS kernel of ART25.}
\label{fig:CS}
\end{figure*}

\textbf{Size of KPCs in various kinematic limits.} To start with, we consider the relative difference between the reduced quasi-TMD correlators computed with KPCs (\ref{F=KPC}) and in the LP approximation (\ref{F=LP}). This comparison is shown in figure~\ref{fig:RelDiff} as the relative deviation between $F^{\text{KPC}}$ and $F^{\text{LP}}$ for different cases. Let us summarize our observations in the following points:
\begin{itemize}
\item We observe that the size of KPC drops very fast with the increase of $P_v$, and it is not that large compared to the contribution of KPCs in the Drell-Yan reaction \cite{Piloneta:2024aac} or SIDIS \cite{Piloneta:2025jjb}. At $P_v\sim 10$ GeV, it is generally less than 5\%.
\item The size of the correction critically depends on the values of $x$ and $b$, and it is not sign-definite. For sufficiently large values of $b$ ($b\gtrsim 1.5$ GeV$^{-1}$), the $x$-dependence becomes flatter and demonstrates a plateau in the range $0.3 \lesssim x \lesssim 0.7$. Consequently, this region is the best for practical studies, because within it the power corrections do not significantly modify the shape of the observable, although they could modify the magnitude. It is interesting to mention that lattice studies also found that the determination of the CS kernel depends on the value of $x$, and is stable in a similar regime \cite{Bollweg:2024zet, Bollweg:2025iol, Avkhadiev:2024mgd, LatticePartonCollaborationLPC:2022myp}, which practically confirms our observation.
\item In this region, the correction is negative. This implies that the lattice extractions of TMD distributions (based on the LP approximation) underestimate its absolute value.
\item The difference between the $\gamma^0$ and $\gamma^v$ cases is almost negligible in comparison to the size of the deviation between KPC and LP, and generally of the order of 5\%. This observation perfectly coincides with lattice simulations (see, e.g., figure 17 in \cite{LatticePartonCollaborationLPC:2022myp}).
\item The size of the correction depends significantly on $b$. For $P_v=1.7$ GeV, which is a typical scale of modern lattice simulations, the correction ranges from $-20\%$ (at $b=3$ GeV$^{-1}$) to $-5\%$ (at $b=5$ GeV$^{-1}$).
\end{itemize}

The general conclusion that can be drawn from this comparison is that, at the present momentum scales of lattice simulations, the LP factorization is not precise and possesses an error of the order of $10$–$20\%$ due to KPCs.

\textbf{Impact to CS kernel.} Due to the intricate relation between quasi-TMDs and TMDs in the TMD-with-KPC factorization approach (equation (\ref{main})), a direct computation of the impact to the extraction of nonperturbative functions is not possible. To estimate the impact of KPCs, we adopt the following procedure. First, we use the ART25 phenomenological input to compute $F^{\text{KPC}}$ (\ref{F=KPC}), which serves as a "pseudo-measurement" based on a known input. Next, we use the LP expression (\ref{F=LP}) to extract the target observable -- either the CS kernel or a TMDPDF -- following the standard strategy employed in modern lattice studies \cite{Bollweg:2024zet, Bollweg:2025iol, Avkhadiev:2024mgd}. This procedure produces a "lattice determination" of corresponding TMD distributions. By comparing this "lattice determinations" with the original input, we can quantify the impact of the missing power corrections.

The determination of CS kernel is the simplest application of quasi-TMDs. At LP it could be extracted considering the quotient of reduced quasi-TMDs evaluated at different values of the hadron momentum \cite{Ebert:2019okf, Ji:2019ewn, Vladimirov:2020ofp}. Explicitly, the expression is
\begin{eqnarray}\label{CS}
&&\mathcal{D}_{\text{LP-like}}(b,\mu)=\frac{1}{2\log(P_{v,2}/P_{v,1})}\,
\\\nn && \qquad\log\(\frac{
C\(2xP_{v,2}/\mu\)F^{[\Gamma],\text{KPCs}}_{q/h}(x,b;\mu;P_{v,1})}{C\(2xP_{v,1}/\mu\)F^{[\Gamma],\text{KPCs}}_{q/h}(x,b;\mu;P_{v,2})}\)\,.
\end{eqnarray}
where $P_{v,i}=v\cdot P_i$. This is currently the common method employed for the extraction of CS kernel from lattice data due to the fact that the reduced soft factors cancel in the ratio, as well as, the actual TMDPDF. Note, that although it is not indicated in the argument, the value of $\mathcal{D}_{\text{LP-like}}$ depends implicitly on $x$, $P_1$ and $P_2$. 

The value of $\mathcal{D}_{\text{LP-like}}$ can be interpreted as the CS kernel that would be extracted from a lattice simulation using the LP formalism, given a specific input CS kernel and TMD distributions. Clearly, this value does not coincide with the actual CS kernel, because $\mathcal{D}_{\text{LP-like}}$ is correlated with the values of the TMDPDFs. The comparison of $\mathcal{D}_{\text{LP-like}}$ with the input $\mathcal{D}$ is shown in figures~\ref{fig:CSxdependence} and \ref{fig:CS} for several kinematical configurations.

In fig.~\ref{fig:CSxdependence}, at fixed $b=2$ GeV$^{-1}$, we present the $x$-dependence of $\mathcal{D}_{\text{LP-like}}$. As discussed earlier, it exhibits a plateau for $0.3 \lesssim x \lesssim 0.7$. The $b$-dependence of $\mathcal{D}_{\text{LP-like}}$ at fixed $x$ is shown in fig.~\ref{fig:CS}. In both cases, we use momentum values typical of lattice simulations~\cite{Avkhadiev:2024mgd}. The uncertainty band is derived from the ART25 uncertainty, and is larger than the uncertainty band for the CS kernel alone, because it also incorporates the uncertainty of the TMDPDFs.

\begin{figure}[t]
\centering
\includegraphics[width=0.4\textwidth]{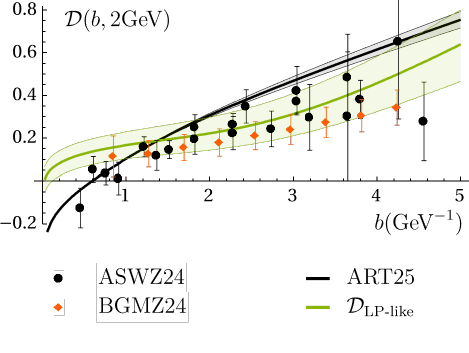}
\caption{CS kernel at $\mu=2$ GeV, obtained by averaging the results of eq.~(\ref{CS}) over $x \in [0.3,0.7]$, $P_v=\{1.29,1.81,2.32\}$, and $\Gamma={\gamma^0,\gamma^v}$ (green band). The black line represents the input CS kernel from ART25. The points show the lattice extractions~\cite{Avkhadiev:2024mgd} and~\cite{Bollweg:2024zet}, performed in a similar kinematic range.}
\label{fig:CSaveraged}
\end{figure}

In fig.~\ref{fig:CSaveraged}, we present $\mathcal{D}_{\text{LP-like}}$ averaged over the region $x \in [0.3,0.7]$, for momentum configurations $P_v = \{1.29, 1.81, 2.32\}$ GeV and $\Gamma = {\gamma^0, \gamma^v}$. The considered kinematic configuration roughly corresponds to that used in ref.~\cite{Avkhadiev:2024mgd}. There is a clear difference between the input ART25 CS kernel and the result of the LP analysis. The resulting $\mathcal{D}_{\text{LP-like}}$ is in very good agreement with the determination of ref.~\cite{Avkhadiev:2024mgd} (we also show the points from ref.~\cite{Bollweg:2024zet} for comparison). Therefore, we conclude that the measurement in ref.~\cite{Avkhadiev:2024mgd} is in agreement with ART25 once KPCs are taken into account.

\begin{figure*}[t]
\centering
\includegraphics[width=0.95\textwidth]{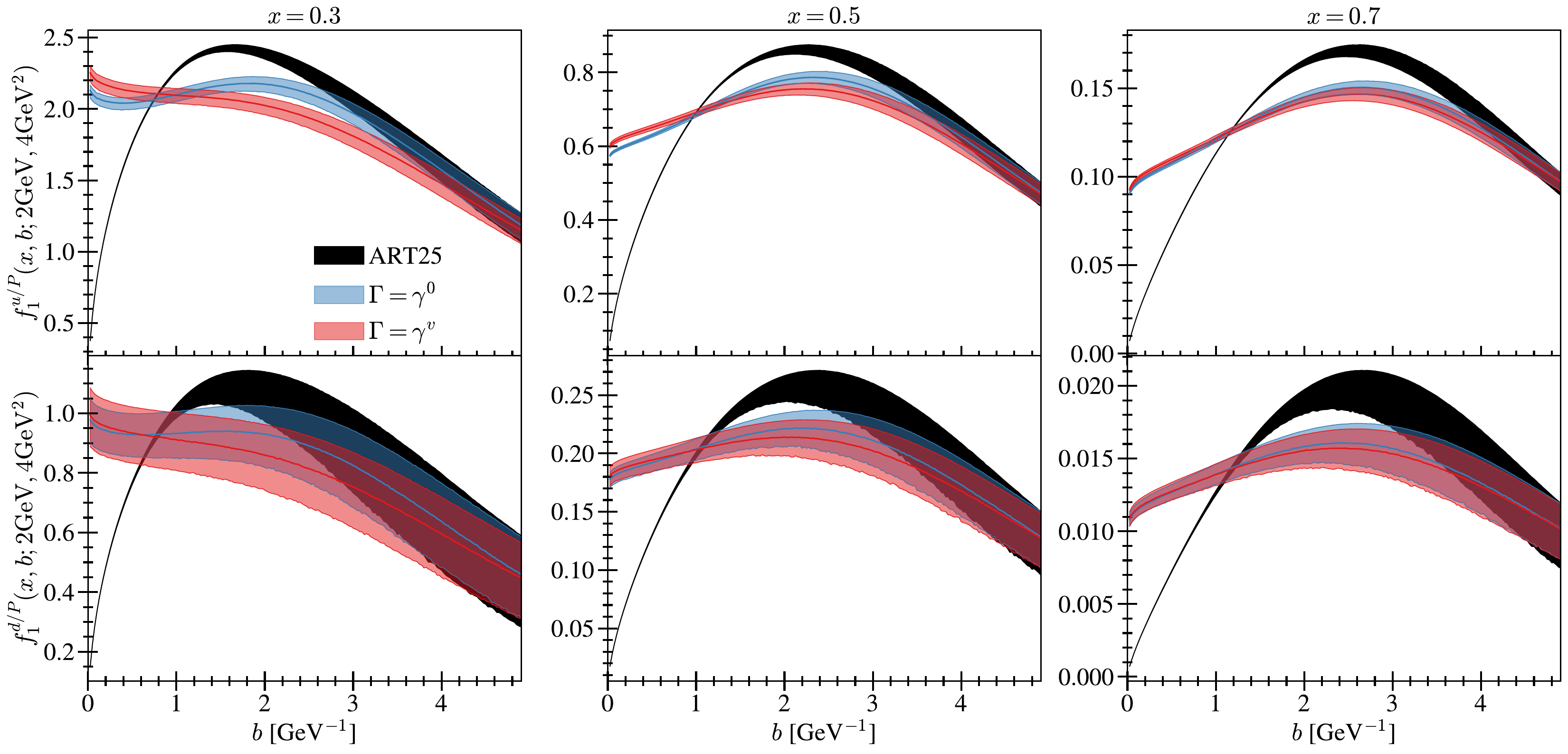}
\caption{Unpolarized TMDPDFs obtained from (\ref{TMDPDF}) for $u$ (upper panels) and $d$ (lower panels) valence quarks inside the proton, as a function of the transverse distance, for three fixed values of the momentum fraction: $x=0.3$ (left), $x=0.5$ (center), and $x=0.7$ (right panels), and for $\Gamma=\gamma^0,\gamma^v$. The renormalization scales are taken to be $\mu=\sqrt{\zeta}=2$ GeV. The uncertainty bands are computed by propagating the uncertainty band of ART25.}
\label{fig:protonb}
\end{figure*}

\textbf{Impact for TMDPDFs.} A similar comparison can be made for the actual TMD distributions. For this purpose, we construct a function that corresponds to the TMDPDF determined using the LP formula. Such a function is given by
\begin{eqnarray}\label{TMDPDF}
f_{\text{LP-like}}^{q/h}(x,b;\mu)=\frac{F^{[\Gamma],\,\rm{KPCs}}_{q/h}(x,b;\mu)}{C\(2xP_v/\mu\)}\,.
\end{eqnarray}
Note that here the value of the CS kernel is taken as given, while in an actual lattice computations it is determined from the same calculation, i.e. it corresponds to $\mathcal{D}_{\text{LP-like}}$.

The comparison of $f_{\text{LP-like}}$ with the input TMDPDF is shown in fig.~\ref{fig:protonb} for several kinematical configurations. In general, one can see that the corrections are of the order of 10–20\% for $b \gtrsim 1.5$ GeV$^{-1}$, and the value of the TMDPDF is underestimated in the LP computation. This is the same conclusion as that drawn from fig.~\ref{fig:RelDiff}.

\section{Conclusion}
\label{sec:conclusions}

In this work, we have derived the factorization theorem for the quasi–transverse-momentum-dependent (quasi-TMD) matrix element, improved by the inclusion of kinematic power corrections (KPCs) at all powers (the so-called TMD-with-KPC factorization theorem \cite{Vladimirov:2023aot}). The resulting expression incorporates the complete contribution of twist-two TMD distributions and is frame-invariant. The expression has the same perturbative hard coefficient function as the LP factorization theorems and is valid for all polarization cases.

The KPCs are only a part of the variety of power corrections contributing to the TMD factorization theorem. However, we argue that they represent the most important part, because they restore frame invariance and other continuous symmetries of the original observable that are broken at the LP approximation. The corrections to the TMD-with-KPC factorization theorem incorporate higher-twist TMD distributions, but do not include twist-two distributions (apart from unknown target-mass contributions). In that sense, the TMD-with-KPC factorization theorem can be considered as the proper form of the LP factorization theorem. In a nutshell, the main difference between the LP and KPC approaches is that the latter takes into account the non-trivial parton kinematics with non-zero transverse momentum.

The improved version of the factorization theorem is presented in (\ref{main}). This expression is valid for any Dirac structure that contributes at LP, and also for any polarization of the hadron. As in the LP factorization, the reduced soft factor is factorized multiplicatively. Due to this, quasi-TMD correlators remain clean observables for TMD physics, since the reduced soft factor can be computed separately and eliminated from the quasi-TMD correlator at all powers.

The crucial difference between the LP and TMD-with-KPC expressions is that the latter relates the values of the parton’s $x$ and $k_T$. As a consequence, the evolution factor cannot be multiplicatively factorized, but enters in an integral convolution. This prevents a direct determination of the Collins–Soper (CS) kernel from the ratio of quasi-TMD correlators, as is usually assumed. Instead, nonperturbative elements should be extracted from lattice data by other procedures, such as from other TMD observables.

We have estimated the size of KPCs and found that it crucially depends on the momentum of the hadron. In particular, they can be considered very small for $P_v \gtrsim 10$ GeV. For the actual values of $P_v$ (of order 1–2 GeV), the corrections are $10$–$20\%$ with respect to the LP factorization theorem. The corrections grow significantly for small and large values of $x$, and possess a plateau at $0.3 \lesssim x \lesssim 0.7$. These corrections translate directly into the TMDPDF; however, they are larger for the CS kernel, which is a more delicate observable. The inclusion of KPCs can change the interpretation of already known results. In particular, we find very good agreement between the ART25 extraction and the results of lattice simulations within the updated approach, although the LP comparison indicates a disagreement.

\textbf{Acknowledgments.} This work is supported by the Grant No. PID2022-136510NB-C31 funded by MCIN/AEI/ 10.13039/501100011033 by the Spanish Ministerio de Ciencias y Innovación. A.V. is supported by the Atracción de Talento Investigador program of the Comunidad de Madrid (Spain) No. 2020-T1/TIC-20204

\bibliography{bibFILE}

\begin{thebibliography}{61}%
\makeatletter
\providecommand \@ifxundefined [1]{%
 \@ifx{#1\undefined}
}%
\providecommand \@ifnum [1]{%
 \ifnum #1\expandafter \@firstoftwo
 \else \expandafter \@secondoftwo
 \fi
}%
\providecommand \@ifx [1]{%
 \ifx #1\expandafter \@firstoftwo
 \else \expandafter \@secondoftwo
 \fi
}%
\providecommand \natexlab [1]{#1}%
\providecommand \enquote  [1]{``#1''}%
\providecommand \bibnamefont  [1]{#1}%
\providecommand \bibfnamefont [1]{#1}%
\providecommand \citenamefont [1]{#1}%
\providecommand \href@noop [0]{\@secondoftwo}%
\providecommand \href [0]{\begingroup \@sanitize@url \@href}%
\providecommand \@href[1]{\@@startlink{#1}\@@href}%
\providecommand \@@href[1]{\endgroup#1\@@endlink}%
\providecommand \@sanitize@url [0]{\catcode `\\12\catcode `\$12\catcode
  `\&12\catcode `\#12\catcode `\^12\catcode `\_12\catcode `\%12\relax}%
\providecommand \@@startlink[1]{}%
\providecommand \@@endlink[0]{}%
\providecommand \url  [0]{\begingroup\@sanitize@url \@url }%
\providecommand \@url [1]{\endgroup\@href {#1}{\urlprefix }}%
\providecommand \urlprefix  [0]{URL }%
\providecommand \Eprint [0]{\href }%
\providecommand \doibase [0]{http://dx.doi.org/}%
\providecommand \selectlanguage [0]{\@gobble}%
\providecommand \bibinfo  [0]{\@secondoftwo}%
\providecommand \bibfield  [0]{\@secondoftwo}%
\providecommand \translation [1]{[#1]}%
\providecommand \BibitemOpen [0]{}%
\providecommand \bibitemStop [0]{}%
\providecommand \bibitemNoStop [0]{.\EOS\space}%
\providecommand \EOS [0]{\spacefactor3000\relax}%
\providecommand \BibitemShut  [1]{\csname bibitem#1\endcsname}%
\let\auto@bib@innerbib\@empty
\bibitem [{\citenamefont {Ji}(2013)}]{Ji:2013dva}%
  \BibitemOpen
  \bibfield  {author} {\bibinfo {author} {\bibfnamefont {X.}~\bibnamefont
  {Ji}},\ }\href {\doibase 10.1103/PhysRevLett.110.262002} {\bibfield
  {journal} {\bibinfo  {journal} {Phys. Rev. Lett.}\ }\textbf {\bibinfo
  {volume} {110}},\ \bibinfo {pages} {262002} (\bibinfo {year} {2013})},\
  \Eprint {http://arxiv.org/abs/1305.1539} {arXiv:1305.1539 [hep-ph]}
  \BibitemShut {NoStop}%
\bibitem [{\citenamefont {Ma}\ and\ \citenamefont {Qiu}(2018)}]{Ma:2014jla}%
  \BibitemOpen
  \bibfield  {author} {\bibinfo {author} {\bibfnamefont {Y.-Q.}\ \bibnamefont
  {Ma}}\ and\ \bibinfo {author} {\bibfnamefont {J.-W.}\ \bibnamefont {Qiu}},\
  }\href {\doibase 10.1103/PhysRevD.98.074021} {\bibfield  {journal} {\bibinfo
  {journal} {Phys. Rev. D}\ }\textbf {\bibinfo {volume} {98}},\ \bibinfo
  {pages} {074021} (\bibinfo {year} {2018})},\ \Eprint
  {http://arxiv.org/abs/1404.6860} {arXiv:1404.6860 [hep-ph]} \BibitemShut
  {NoStop}%
\bibitem [{\citenamefont {Constantinou}\ \emph {et~al.}(2022)\citenamefont
  {Constantinou} \emph {et~al.}}]{Constantinou:2022yye}%
  \BibitemOpen
  \bibfield  {author} {\bibinfo {author} {\bibfnamefont {M.}~\bibnamefont
  {Constantinou}} \emph {et~al.},\ }\href@noop {} {\  (\bibinfo {year}
  {2022})},\ \Eprint {http://arxiv.org/abs/2202.07193} {arXiv:2202.07193
  [hep-lat]} \BibitemShut {NoStop}%
\bibitem [{\citenamefont {Radyushkin}(2017)}]{Radyushkin:2017cyf}%
  \BibitemOpen
  \bibfield  {author} {\bibinfo {author} {\bibfnamefont {A.~V.}\ \bibnamefont
  {Radyushkin}},\ }\href {\doibase 10.1103/PhysRevD.96.034025} {\bibfield
  {journal} {\bibinfo  {journal} {Phys. Rev. D}\ }\textbf {\bibinfo {volume}
  {96}},\ \bibinfo {pages} {034025} (\bibinfo {year} {2017})},\ \Eprint
  {http://arxiv.org/abs/1705.01488} {arXiv:1705.01488 [hep-ph]} \BibitemShut
  {NoStop}%
\bibitem [{\citenamefont {Izubuchi}\ \emph {et~al.}(2018)\citenamefont
  {Izubuchi}, \citenamefont {Ji}, \citenamefont {Jin}, \citenamefont
  {Stewart},\ and\ \citenamefont {Zhao}}]{Izubuchi:2018srq}%
  \BibitemOpen
  \bibfield  {author} {\bibinfo {author} {\bibfnamefont {T.}~\bibnamefont
  {Izubuchi}}, \bibinfo {author} {\bibfnamefont {X.}~\bibnamefont {Ji}},
  \bibinfo {author} {\bibfnamefont {L.}~\bibnamefont {Jin}}, \bibinfo {author}
  {\bibfnamefont {I.~W.}\ \bibnamefont {Stewart}}, \ and\ \bibinfo {author}
  {\bibfnamefont {Y.}~\bibnamefont {Zhao}},\ }\href {\doibase
  10.1103/PhysRevD.98.056004} {\bibfield  {journal} {\bibinfo  {journal} {Phys.
  Rev. D}\ }\textbf {\bibinfo {volume} {98}},\ \bibinfo {pages} {056004}
  (\bibinfo {year} {2018})},\ \Eprint {http://arxiv.org/abs/1801.03917}
  {arXiv:1801.03917 [hep-ph]} \BibitemShut {NoStop}%
\bibitem [{\citenamefont {Ji}\ \emph {et~al.}(2021)\citenamefont {Ji},
  \citenamefont {Liu}, \citenamefont {Liu}, \citenamefont {Zhang},\ and\
  \citenamefont {Zhao}}]{Ji:2020ect}%
  \BibitemOpen
  \bibfield  {author} {\bibinfo {author} {\bibfnamefont {X.}~\bibnamefont
  {Ji}}, \bibinfo {author} {\bibfnamefont {Y.-S.}\ \bibnamefont {Liu}},
  \bibinfo {author} {\bibfnamefont {Y.}~\bibnamefont {Liu}}, \bibinfo {author}
  {\bibfnamefont {J.-H.}\ \bibnamefont {Zhang}}, \ and\ \bibinfo {author}
  {\bibfnamefont {Y.}~\bibnamefont {Zhao}},\ }\href {\doibase
  10.1103/RevModPhys.93.035005} {\bibfield  {journal} {\bibinfo  {journal}
  {Rev. Mod. Phys.}\ }\textbf {\bibinfo {volume} {93}},\ \bibinfo {pages}
  {035005} (\bibinfo {year} {2021})},\ \Eprint
  {http://arxiv.org/abs/2004.03543} {arXiv:2004.03543 [hep-ph]} \BibitemShut
  {NoStop}%
\bibitem [{\citenamefont {Ji}\ \emph {et~al.}(2019)\citenamefont {Ji},
  \citenamefont {Jin}, \citenamefont {Yuan}, \citenamefont {Zhang},\ and\
  \citenamefont {Zhao}}]{Ji:2018hvs}%
  \BibitemOpen
  \bibfield  {author} {\bibinfo {author} {\bibfnamefont {X.}~\bibnamefont
  {Ji}}, \bibinfo {author} {\bibfnamefont {L.-C.}\ \bibnamefont {Jin}},
  \bibinfo {author} {\bibfnamefont {F.}~\bibnamefont {Yuan}}, \bibinfo {author}
  {\bibfnamefont {J.-H.}\ \bibnamefont {Zhang}}, \ and\ \bibinfo {author}
  {\bibfnamefont {Y.}~\bibnamefont {Zhao}},\ }\href {\doibase
  10.1103/PhysRevD.99.114006} {\bibfield  {journal} {\bibinfo  {journal} {Phys.
  Rev. D}\ }\textbf {\bibinfo {volume} {99}},\ \bibinfo {pages} {114006}
  (\bibinfo {year} {2019})},\ \Eprint {http://arxiv.org/abs/1801.05930}
  {arXiv:1801.05930 [hep-ph]} \BibitemShut {NoStop}%
\bibitem [{\citenamefont {Ji}\ \emph {et~al.}(2020)\citenamefont {Ji},
  \citenamefont {Liu},\ and\ \citenamefont {Liu}}]{Ji:2019ewn}%
  \BibitemOpen
  \bibfield  {author} {\bibinfo {author} {\bibfnamefont {X.}~\bibnamefont
  {Ji}}, \bibinfo {author} {\bibfnamefont {Y.}~\bibnamefont {Liu}}, \ and\
  \bibinfo {author} {\bibfnamefont {Y.-S.}\ \bibnamefont {Liu}},\ }\href
  {\doibase 10.1016/j.physletb.2020.135946} {\bibfield  {journal} {\bibinfo
  {journal} {Phys. Lett. B}\ }\textbf {\bibinfo {volume} {811}},\ \bibinfo
  {pages} {135946} (\bibinfo {year} {2020})},\ \Eprint
  {http://arxiv.org/abs/1911.03840} {arXiv:1911.03840 [hep-ph]} \BibitemShut
  {NoStop}%
\bibitem [{\citenamefont {Ebert}\ \emph {et~al.}(2019)\citenamefont {Ebert},
  \citenamefont {Stewart},\ and\ \citenamefont {Zhao}}]{Ebert:2019okf}%
  \BibitemOpen
  \bibfield  {author} {\bibinfo {author} {\bibfnamefont {M.~A.}\ \bibnamefont
  {Ebert}}, \bibinfo {author} {\bibfnamefont {I.~W.}\ \bibnamefont {Stewart}},
  \ and\ \bibinfo {author} {\bibfnamefont {Y.}~\bibnamefont {Zhao}},\ }\href
  {\doibase 10.1007/JHEP09(2019)037} {\bibfield  {journal} {\bibinfo  {journal}
  {JHEP}\ }\textbf {\bibinfo {volume} {09}},\ \bibinfo {pages} {037} (\bibinfo
  {year} {2019})},\ \Eprint {http://arxiv.org/abs/1901.03685} {arXiv:1901.03685
  [hep-ph]} \BibitemShut {NoStop}%
\bibitem [{\citenamefont {Vladimirov}\ and\ \citenamefont
  {Sch\"afer}(2020)}]{Vladimirov:2020ofp}%
  \BibitemOpen
  \bibfield  {author} {\bibinfo {author} {\bibfnamefont {A.~A.}\ \bibnamefont
  {Vladimirov}}\ and\ \bibinfo {author} {\bibfnamefont {A.}~\bibnamefont
  {Sch\"afer}},\ }\href {\doibase 10.1103/PhysRevD.101.074517} {\bibfield
  {journal} {\bibinfo  {journal} {Phys. Rev. D}\ }\textbf {\bibinfo {volume}
  {101}},\ \bibinfo {pages} {074517} (\bibinfo {year} {2020})},\ \Eprint
  {http://arxiv.org/abs/2002.07527} {arXiv:2002.07527 [hep-ph]} \BibitemShut
  {NoStop}%
\bibitem [{\citenamefont {Ebert}\ \emph {et~al.}(2022)\citenamefont {Ebert},
  \citenamefont {Schindler}, \citenamefont {Stewart},\ and\ \citenamefont
  {Zhao}}]{Ebert:2022fmh}%
  \BibitemOpen
  \bibfield  {author} {\bibinfo {author} {\bibfnamefont {M.~A.}\ \bibnamefont
  {Ebert}}, \bibinfo {author} {\bibfnamefont {S.~T.}\ \bibnamefont
  {Schindler}}, \bibinfo {author} {\bibfnamefont {I.~W.}\ \bibnamefont
  {Stewart}}, \ and\ \bibinfo {author} {\bibfnamefont {Y.}~\bibnamefont
  {Zhao}},\ }\href {\doibase 10.1007/JHEP04(2022)178} {\bibfield  {journal}
  {\bibinfo  {journal} {JHEP}\ }\textbf {\bibinfo {volume} {04}},\ \bibinfo
  {pages} {178} (\bibinfo {year} {2022})},\ \Eprint
  {http://arxiv.org/abs/2201.08401} {arXiv:2201.08401 [hep-ph]} \BibitemShut
  {NoStop}%
\bibitem [{\citenamefont {Constantinou}\ \emph {et~al.}(2021)\citenamefont
  {Constantinou} \emph {et~al.}}]{Constantinou:2020hdm}%
  \BibitemOpen
  \bibfield  {author} {\bibinfo {author} {\bibfnamefont {M.}~\bibnamefont
  {Constantinou}} \emph {et~al.},\ }\href {\doibase 10.1016/j.ppnp.2021.103908}
  {\bibfield  {journal} {\bibinfo  {journal} {Prog. Part. Nucl. Phys.}\
  }\textbf {\bibinfo {volume} {121}},\ \bibinfo {pages} {103908} (\bibinfo
  {year} {2021})},\ \Eprint {http://arxiv.org/abs/2006.08636} {arXiv:2006.08636
  [hep-ph]} \BibitemShut {NoStop}%
\bibitem [{\citenamefont {Bhattacharya}\ \emph {et~al.}(2022)\citenamefont
  {Bhattacharya}, \citenamefont {Cichy}, \citenamefont {Constantinou},
  \citenamefont {Dodson}, \citenamefont {Gao}, \citenamefont {Metz},
  \citenamefont {Mukherjee}, \citenamefont {Scapellato}, \citenamefont
  {Steffens},\ and\ \citenamefont {Zhao}}]{Bhattacharya:2022aob}%
  \BibitemOpen
  \bibfield  {author} {\bibinfo {author} {\bibfnamefont {S.}~\bibnamefont
  {Bhattacharya}}, \bibinfo {author} {\bibfnamefont {K.}~\bibnamefont {Cichy}},
  \bibinfo {author} {\bibfnamefont {M.}~\bibnamefont {Constantinou}}, \bibinfo
  {author} {\bibfnamefont {J.}~\bibnamefont {Dodson}}, \bibinfo {author}
  {\bibfnamefont {X.}~\bibnamefont {Gao}}, \bibinfo {author} {\bibfnamefont
  {A.}~\bibnamefont {Metz}}, \bibinfo {author} {\bibfnamefont {S.}~\bibnamefont
  {Mukherjee}}, \bibinfo {author} {\bibfnamefont {A.}~\bibnamefont
  {Scapellato}}, \bibinfo {author} {\bibfnamefont {F.}~\bibnamefont
  {Steffens}}, \ and\ \bibinfo {author} {\bibfnamefont {Y.}~\bibnamefont
  {Zhao}},\ }\href {\doibase 10.1103/PhysRevD.106.114512} {\bibfield  {journal}
  {\bibinfo  {journal} {Phys. Rev. D}\ }\textbf {\bibinfo {volume} {106}},\
  \bibinfo {pages} {114512} (\bibinfo {year} {2022})},\ \Eprint
  {http://arxiv.org/abs/2209.05373} {arXiv:2209.05373 [hep-lat]} \BibitemShut
  {NoStop}%
\bibitem [{\citenamefont {Bali}\ \emph {et~al.}(2016)\citenamefont {Bali},
  \citenamefont {Lang}, \citenamefont {Musch},\ and\ \citenamefont
  {Sch{\"a}fer}}]{Bali:2016lva}%
  \BibitemOpen
  \bibfield  {author} {\bibinfo {author} {\bibfnamefont {G.~S.}\ \bibnamefont
  {Bali}}, \bibinfo {author} {\bibfnamefont {B.}~\bibnamefont {Lang}}, \bibinfo
  {author} {\bibfnamefont {B.~U.}\ \bibnamefont {Musch}}, \ and\ \bibinfo
  {author} {\bibfnamefont {A.}~\bibnamefont {Sch{\"a}fer}},\ }\href {\doibase
  10.1103/PhysRevD.93.094515} {\bibfield  {journal} {\bibinfo  {journal} {Phys.
  Rev. D}\ }\textbf {\bibinfo {volume} {93}},\ \bibinfo {pages} {094515}
  (\bibinfo {year} {2016})},\ \Eprint {http://arxiv.org/abs/1602.05525}
  {arXiv:1602.05525 [hep-lat]} \BibitemShut {NoStop}%
\bibitem [{\citenamefont {Angeles-Martinez}\ \emph {et~al.}(2015)\citenamefont
  {Angeles-Martinez} \emph {et~al.}}]{Angeles-Martinez:2015sea}%
  \BibitemOpen
  \bibfield  {author} {\bibinfo {author} {\bibfnamefont {R.}~\bibnamefont
  {Angeles-Martinez}} \emph {et~al.},\ }\href {\doibase
  10.5506/APhysPolB.46.2501} {\bibfield  {journal} {\bibinfo  {journal} {Acta
  Phys. Polon. B}\ }\textbf {\bibinfo {volume} {46}},\ \bibinfo {pages} {2501}
  (\bibinfo {year} {2015})},\ \Eprint {http://arxiv.org/abs/1507.05267}
  {arXiv:1507.05267 [hep-ph]} \BibitemShut {NoStop}%
\bibitem [{\citenamefont {Boussarie}\ \emph {et~al.}(2023)\citenamefont
  {Boussarie} \emph {et~al.}}]{Boussarie:2023izj}%
  \BibitemOpen
  \bibfield  {author} {\bibinfo {author} {\bibfnamefont {R.}~\bibnamefont
  {Boussarie}} \emph {et~al.},\ }\href@noop {} {\  (\bibinfo {year} {2023})},\
  \Eprint {http://arxiv.org/abs/2304.03302} {arXiv:2304.03302 [hep-ph]}
  \BibitemShut {NoStop}%
\bibitem [{\citenamefont {Rodini}\ and\ \citenamefont
  {Vladimirov}(2023)}]{Rodini:2022wic}%
  \BibitemOpen
  \bibfield  {author} {\bibinfo {author} {\bibfnamefont {S.}~\bibnamefont
  {Rodini}}\ and\ \bibinfo {author} {\bibfnamefont {A.}~\bibnamefont
  {Vladimirov}},\ }\href {\doibase 10.1007/JHEP09(2023)117} {\bibfield
  {journal} {\bibinfo  {journal} {JHEP}\ }\textbf {\bibinfo {volume} {09}},\
  \bibinfo {pages} {117} (\bibinfo {year} {2023})},\ \Eprint
  {http://arxiv.org/abs/2211.04494} {arXiv:2211.04494 [hep-ph]} \BibitemShut
  {NoStop}%
\bibitem [{\citenamefont {Zhang}\ \emph {et~al.}(2020)\citenamefont {Zhang}
  \emph {et~al.}}]{LatticeParton:2020uhz}%
  \BibitemOpen
  \bibfield  {author} {\bibinfo {author} {\bibfnamefont {Q.-A.}\ \bibnamefont
  {Zhang}} \emph {et~al.} (\bibinfo {collaboration} {Lattice Parton}),\ }\href
  {\doibase 10.22323/1.396.0477} {\bibfield  {journal} {\bibinfo  {journal}
  {Phys. Rev. Lett.}\ }\textbf {\bibinfo {volume} {125}},\ \bibinfo {pages}
  {192001} (\bibinfo {year} {2020})},\ \Eprint
  {http://arxiv.org/abs/2005.14572} {arXiv:2005.14572 [hep-lat]} \BibitemShut
  {NoStop}%
\bibitem [{\citenamefont {Schlemmer}\ \emph {et~al.}(2021)\citenamefont
  {Schlemmer}, \citenamefont {Vladimirov}, \citenamefont {Zimmermann},
  \citenamefont {Engelhardt},\ and\ \citenamefont
  {Sch\"afer}}]{Schlemmer:2021aij}%
  \BibitemOpen
  \bibfield  {author} {\bibinfo {author} {\bibfnamefont {M.}~\bibnamefont
  {Schlemmer}}, \bibinfo {author} {\bibfnamefont {A.}~\bibnamefont
  {Vladimirov}}, \bibinfo {author} {\bibfnamefont {C.}~\bibnamefont
  {Zimmermann}}, \bibinfo {author} {\bibfnamefont {M.}~\bibnamefont
  {Engelhardt}}, \ and\ \bibinfo {author} {\bibfnamefont {A.}~\bibnamefont
  {Sch\"afer}},\ }\href {\doibase 10.1007/JHEP08(2021)004} {\bibfield
  {journal} {\bibinfo  {journal} {JHEP}\ }\textbf {\bibinfo {volume} {08}},\
  \bibinfo {pages} {004} (\bibinfo {year} {2021})},\ \Eprint
  {http://arxiv.org/abs/2103.16991} {arXiv:2103.16991 [hep-lat]} \BibitemShut
  {NoStop}%
\bibitem [{\citenamefont {Shanahan}\ \emph {et~al.}(2021)\citenamefont
  {Shanahan}, \citenamefont {Wagman},\ and\ \citenamefont
  {Zhao}}]{Shanahan:2021tst}%
  \BibitemOpen
  \bibfield  {author} {\bibinfo {author} {\bibfnamefont {P.}~\bibnamefont
  {Shanahan}}, \bibinfo {author} {\bibfnamefont {M.}~\bibnamefont {Wagman}}, \
  and\ \bibinfo {author} {\bibfnamefont {Y.}~\bibnamefont {Zhao}},\ }\href
  {\doibase 10.1103/PhysRevD.104.114502} {\bibfield  {journal} {\bibinfo
  {journal} {Phys. Rev. D}\ }\textbf {\bibinfo {volume} {104}},\ \bibinfo
  {pages} {114502} (\bibinfo {year} {2021})},\ \Eprint
  {http://arxiv.org/abs/2107.11930} {arXiv:2107.11930 [hep-lat]} \BibitemShut
  {NoStop}%
\bibitem [{\citenamefont {Li}\ \emph {et~al.}(2022)\citenamefont {Li} \emph
  {et~al.}}]{Li:2021wvl}%
  \BibitemOpen
  \bibfield  {author} {\bibinfo {author} {\bibfnamefont {Y.}~\bibnamefont {Li}}
  \emph {et~al.},\ }\href {\doibase 10.1103/PhysRevLett.128.062002} {\bibfield
  {journal} {\bibinfo  {journal} {Phys. Rev. Lett.}\ }\textbf {\bibinfo
  {volume} {128}},\ \bibinfo {pages} {062002} (\bibinfo {year} {2022})},\
  \Eprint {http://arxiv.org/abs/2106.13027} {arXiv:2106.13027 [hep-lat]}
  \BibitemShut {NoStop}%
\bibitem [{\citenamefont {Chu}\ \emph {et~al.}(2022)\citenamefont {Chu} \emph
  {et~al.}}]{LatticePartonLPC:2022eev}%
  \BibitemOpen
  \bibfield  {author} {\bibinfo {author} {\bibfnamefont {M.-H.}\ \bibnamefont
  {Chu}} \emph {et~al.} (\bibinfo {collaboration} {Lattice Parton (LPC)}),\
  }\href {\doibase 10.1103/PhysRevD.106.034509} {\bibfield  {journal} {\bibinfo
   {journal} {Phys. Rev. D}\ }\textbf {\bibinfo {volume} {106}},\ \bibinfo
  {pages} {034509} (\bibinfo {year} {2022})},\ \Eprint
  {http://arxiv.org/abs/2204.00200} {arXiv:2204.00200 [hep-lat]} \BibitemShut
  {NoStop}%
\bibitem [{\citenamefont {Shu}\ \emph {et~al.}(2023)\citenamefont {Shu},
  \citenamefont {Schlemmer}, \citenamefont {Sizmann}, \citenamefont
  {Vladimirov}, \citenamefont {Walter}, \citenamefont {Engelhardt},
  \citenamefont {Sch\"afer},\ and\ \citenamefont {Yang}}]{Shu:2023cot}%
  \BibitemOpen
  \bibfield  {author} {\bibinfo {author} {\bibfnamefont {H.-T.}\ \bibnamefont
  {Shu}}, \bibinfo {author} {\bibfnamefont {M.}~\bibnamefont {Schlemmer}},
  \bibinfo {author} {\bibfnamefont {T.}~\bibnamefont {Sizmann}}, \bibinfo
  {author} {\bibfnamefont {A.}~\bibnamefont {Vladimirov}}, \bibinfo {author}
  {\bibfnamefont {L.}~\bibnamefont {Walter}}, \bibinfo {author} {\bibfnamefont
  {M.}~\bibnamefont {Engelhardt}}, \bibinfo {author} {\bibfnamefont
  {A.}~\bibnamefont {Sch\"afer}}, \ and\ \bibinfo {author} {\bibfnamefont
  {Y.-B.}\ \bibnamefont {Yang}},\ }\href {\doibase 10.1103/PhysRevD.108.074519}
  {\bibfield  {journal} {\bibinfo  {journal} {Phys. Rev. D}\ }\textbf {\bibinfo
  {volume} {108}},\ \bibinfo {pages} {074519} (\bibinfo {year} {2023})},\
  \Eprint {http://arxiv.org/abs/2302.06502} {arXiv:2302.06502 [hep-lat]}
  \BibitemShut {NoStop}%
\bibitem [{\citenamefont {Avkhadiev}\ \emph {et~al.}(2023)\citenamefont
  {Avkhadiev}, \citenamefont {Shanahan}, \citenamefont {Wagman},\ and\
  \citenamefont {Zhao}}]{Avkhadiev:2023poz}%
  \BibitemOpen
  \bibfield  {author} {\bibinfo {author} {\bibfnamefont {A.}~\bibnamefont
  {Avkhadiev}}, \bibinfo {author} {\bibfnamefont {P.~E.}\ \bibnamefont
  {Shanahan}}, \bibinfo {author} {\bibfnamefont {M.~L.}\ \bibnamefont
  {Wagman}}, \ and\ \bibinfo {author} {\bibfnamefont {Y.}~\bibnamefont
  {Zhao}},\ }\href {\doibase 10.1103/PhysRevD.108.114505} {\bibfield  {journal}
  {\bibinfo  {journal} {Phys. Rev. D}\ }\textbf {\bibinfo {volume} {108}},\
  \bibinfo {pages} {114505} (\bibinfo {year} {2023})},\ \Eprint
  {http://arxiv.org/abs/2307.12359} {arXiv:2307.12359 [hep-lat]} \BibitemShut
  {NoStop}%
\bibitem [{\citenamefont {Avkhadiev}\ \emph {et~al.}(2024)\citenamefont
  {Avkhadiev}, \citenamefont {Shanahan}, \citenamefont {Wagman},\ and\
  \citenamefont {Zhao}}]{Avkhadiev:2024mgd}%
  \BibitemOpen
  \bibfield  {author} {\bibinfo {author} {\bibfnamefont {A.}~\bibnamefont
  {Avkhadiev}}, \bibinfo {author} {\bibfnamefont {P.~E.}\ \bibnamefont
  {Shanahan}}, \bibinfo {author} {\bibfnamefont {M.~L.}\ \bibnamefont
  {Wagman}}, \ and\ \bibinfo {author} {\bibfnamefont {Y.}~\bibnamefont
  {Zhao}},\ }\href {\doibase 10.1103/PhysRevLett.132.231901} {\bibfield
  {journal} {\bibinfo  {journal} {Phys. Rev. Lett.}\ }\textbf {\bibinfo
  {volume} {132}},\ \bibinfo {pages} {231901} (\bibinfo {year} {2024})},\
  \Eprint {http://arxiv.org/abs/2402.06725} {arXiv:2402.06725 [hep-lat]}
  \BibitemShut {NoStop}%
\bibitem [{\citenamefont {Bollweg}\ \emph {et~al.}(2024)\citenamefont
  {Bollweg}, \citenamefont {Gao}, \citenamefont {Mukherjee},\ and\
  \citenamefont {Zhao}}]{Bollweg:2024zet}%
  \BibitemOpen
  \bibfield  {author} {\bibinfo {author} {\bibfnamefont {D.}~\bibnamefont
  {Bollweg}}, \bibinfo {author} {\bibfnamefont {X.}~\bibnamefont {Gao}},
  \bibinfo {author} {\bibfnamefont {S.}~\bibnamefont {Mukherjee}}, \ and\
  \bibinfo {author} {\bibfnamefont {Y.}~\bibnamefont {Zhao}},\ }\href {\doibase
  10.1016/j.physletb.2024.138617} {\bibfield  {journal} {\bibinfo  {journal}
  {Phys. Lett. B}\ }\textbf {\bibinfo {volume} {852}},\ \bibinfo {pages}
  {138617} (\bibinfo {year} {2024})},\ \Eprint
  {http://arxiv.org/abs/2403.00664} {arXiv:2403.00664 [hep-lat]} \BibitemShut
  {NoStop}%
\bibitem [{\citenamefont {Alexandrou}\ \emph {et~al.}(2025)\citenamefont
  {Alexandrou}, \citenamefont {Bacchio}, \citenamefont {Cichy}, \citenamefont
  {Constantinou}, \citenamefont {Sen}, \citenamefont {Spanoudes}, \citenamefont
  {Steffens},\ and\ \citenamefont {Tarello}}]{Alexandrou:2025xci}%
  \BibitemOpen
  \bibfield  {author} {\bibinfo {author} {\bibfnamefont {C.}~\bibnamefont
  {Alexandrou}}, \bibinfo {author} {\bibfnamefont {S.}~\bibnamefont {Bacchio}},
  \bibinfo {author} {\bibfnamefont {K.}~\bibnamefont {Cichy}}, \bibinfo
  {author} {\bibfnamefont {M.}~\bibnamefont {Constantinou}}, \bibinfo {author}
  {\bibfnamefont {A.}~\bibnamefont {Sen}}, \bibinfo {author} {\bibfnamefont
  {G.}~\bibnamefont {Spanoudes}}, \bibinfo {author} {\bibfnamefont
  {F.}~\bibnamefont {Steffens}}, \ and\ \bibinfo {author} {\bibfnamefont
  {J.}~\bibnamefont {Tarello}},\ }\href@noop {} {\  (\bibinfo {year} {2025})},\
  \Eprint {http://arxiv.org/abs/2509.26316} {arXiv:2509.26316 [hep-lat]}
  \BibitemShut {NoStop}%
\bibitem [{\citenamefont {Tan}\ \emph {et~al.}(2025)\citenamefont {Tan} \emph
  {et~al.}}]{Tan:2025ofx}%
  \BibitemOpen
  \bibfield  {author} {\bibinfo {author} {\bibfnamefont {J.-X.}\ \bibnamefont
  {Tan}} \emph {et~al.},\ }\href@noop {} {\  (\bibinfo {year} {2025})},\
  \Eprint {http://arxiv.org/abs/2511.22547} {arXiv:2511.22547 [hep-lat]}
  \BibitemShut {NoStop}%
\bibitem [{\citenamefont {He}\ \emph {et~al.}(2024)\citenamefont {He},
  \citenamefont {Chu}, \citenamefont {Hua}, \citenamefont {Ji}, \citenamefont
  {Sch{\"a}fer}, \citenamefont {Su}, \citenamefont {Wang}, \citenamefont
  {Yang}, \citenamefont {Zhang},\ and\ \citenamefont
  {Zhang}}]{LatticePartonCollaborationLPC:2022myp}%
  \BibitemOpen
  \bibfield  {author} {\bibinfo {author} {\bibfnamefont {J.-C.}\ \bibnamefont
  {He}}, \bibinfo {author} {\bibfnamefont {M.-H.}\ \bibnamefont {Chu}},
  \bibinfo {author} {\bibfnamefont {J.}~\bibnamefont {Hua}}, \bibinfo {author}
  {\bibfnamefont {X.}~\bibnamefont {Ji}}, \bibinfo {author} {\bibfnamefont
  {A.}~\bibnamefont {Sch{\"a}fer}}, \bibinfo {author} {\bibfnamefont
  {Y.}~\bibnamefont {Su}}, \bibinfo {author} {\bibfnamefont {W.}~\bibnamefont
  {Wang}}, \bibinfo {author} {\bibfnamefont {Y.-B.}\ \bibnamefont {Yang}},
  \bibinfo {author} {\bibfnamefont {J.-H.}\ \bibnamefont {Zhang}}, \ and\
  \bibinfo {author} {\bibfnamefont {Q.-A.}\ \bibnamefont {Zhang}} (\bibinfo
  {collaboration} {Lattice Parton Collaboration (LPC)}),\ }\href {\doibase
  10.1103/PhysRevD.109.114513} {\bibfield  {journal} {\bibinfo  {journal}
  {Phys. Rev. D}\ }\textbf {\bibinfo {volume} {109}},\ \bibinfo {pages}
  {114513} (\bibinfo {year} {2024})},\ \Eprint
  {http://arxiv.org/abs/2211.02340} {arXiv:2211.02340 [hep-lat]} \BibitemShut
  {NoStop}%
\bibitem [{\citenamefont {Bollweg}\ \emph {et~al.}(2025)\citenamefont
  {Bollweg}, \citenamefont {Gao}, \citenamefont {He}, \citenamefont
  {Mukherjee},\ and\ \citenamefont {Zhao}}]{Bollweg:2025iol}%
  \BibitemOpen
  \bibfield  {author} {\bibinfo {author} {\bibfnamefont {D.}~\bibnamefont
  {Bollweg}}, \bibinfo {author} {\bibfnamefont {X.}~\bibnamefont {Gao}},
  \bibinfo {author} {\bibfnamefont {J.}~\bibnamefont {He}}, \bibinfo {author}
  {\bibfnamefont {S.}~\bibnamefont {Mukherjee}}, \ and\ \bibinfo {author}
  {\bibfnamefont {Y.}~\bibnamefont {Zhao}},\ }\href {\doibase
  10.1103/j3n6-8kxy} {\bibfield  {journal} {\bibinfo  {journal} {Phys. Rev. D}\
  }\textbf {\bibinfo {volume} {112}},\ \bibinfo {pages} {034501} (\bibinfo
  {year} {2025})},\ \Eprint {http://arxiv.org/abs/2504.04625} {arXiv:2504.04625
  [hep-lat]} \BibitemShut {NoStop}%
\bibitem [{\citenamefont {Bacchetta}\ \emph {et~al.}(2017)\citenamefont
  {Bacchetta}, \citenamefont {Delcarro}, \citenamefont {Pisano}, \citenamefont
  {Radici},\ and\ \citenamefont {Signori}}]{Bacchetta:2017gcc}%
  \BibitemOpen
  \bibfield  {author} {\bibinfo {author} {\bibfnamefont {A.}~\bibnamefont
  {Bacchetta}}, \bibinfo {author} {\bibfnamefont {F.}~\bibnamefont {Delcarro}},
  \bibinfo {author} {\bibfnamefont {C.}~\bibnamefont {Pisano}}, \bibinfo
  {author} {\bibfnamefont {M.}~\bibnamefont {Radici}}, \ and\ \bibinfo {author}
  {\bibfnamefont {A.}~\bibnamefont {Signori}},\ }\href {\doibase
  10.1007/JHEP06(2017)081} {\bibfield  {journal} {\bibinfo  {journal} {JHEP}\
  }\textbf {\bibinfo {volume} {06}},\ \bibinfo {pages} {081} (\bibinfo {year}
  {2017})},\ \bibinfo {note} {[Erratum: JHEP 06, 051 (2019)]},\ \Eprint
  {http://arxiv.org/abs/1703.10157} {arXiv:1703.10157 [hep-ph]} \BibitemShut
  {NoStop}%
\bibitem [{\citenamefont {Scimemi}\ and\ \citenamefont
  {Vladimirov}(2018)}]{Scimemi:2017etj}%
  \BibitemOpen
  \bibfield  {author} {\bibinfo {author} {\bibfnamefont {I.}~\bibnamefont
  {Scimemi}}\ and\ \bibinfo {author} {\bibfnamefont {A.}~\bibnamefont
  {Vladimirov}},\ }\href {\doibase 10.1140/epjc/s10052-018-5557-y} {\bibfield
  {journal} {\bibinfo  {journal} {Eur. Phys. J. C}\ }\textbf {\bibinfo {volume}
  {78}},\ \bibinfo {pages} {89} (\bibinfo {year} {2018})},\ \Eprint
  {http://arxiv.org/abs/1706.01473} {arXiv:1706.01473 [hep-ph]} \BibitemShut
  {NoStop}%
\bibitem [{\citenamefont {Bertone}\ \emph {et~al.}(2019)\citenamefont
  {Bertone}, \citenamefont {Scimemi},\ and\ \citenamefont
  {Vladimirov}}]{Bertone:2019nxa}%
  \BibitemOpen
  \bibfield  {author} {\bibinfo {author} {\bibfnamefont {V.}~\bibnamefont
  {Bertone}}, \bibinfo {author} {\bibfnamefont {I.}~\bibnamefont {Scimemi}}, \
  and\ \bibinfo {author} {\bibfnamefont {A.}~\bibnamefont {Vladimirov}},\
  }\href {\doibase 10.1007/JHEP06(2019)028} {\bibfield  {journal} {\bibinfo
  {journal} {JHEP}\ }\textbf {\bibinfo {volume} {06}},\ \bibinfo {pages} {028}
  (\bibinfo {year} {2019})},\ \Eprint {http://arxiv.org/abs/1902.08474}
  {arXiv:1902.08474 [hep-ph]} \BibitemShut {NoStop}%
\bibitem [{\citenamefont {Scimemi}\ and\ \citenamefont
  {Vladimirov}(2020)}]{Scimemi:2019cmh}%
  \BibitemOpen
  \bibfield  {author} {\bibinfo {author} {\bibfnamefont {I.}~\bibnamefont
  {Scimemi}}\ and\ \bibinfo {author} {\bibfnamefont {A.}~\bibnamefont
  {Vladimirov}},\ }\href {\doibase 10.1007/JHEP06(2020)137} {\bibfield
  {journal} {\bibinfo  {journal} {JHEP}\ }\textbf {\bibinfo {volume} {06}},\
  \bibinfo {pages} {137} (\bibinfo {year} {2020})},\ \Eprint
  {http://arxiv.org/abs/1912.06532} {arXiv:1912.06532 [hep-ph]} \BibitemShut
  {NoStop}%
\bibitem [{\citenamefont {Vladimirov}(2019)}]{Vladimirov:2019bfa}%
  \BibitemOpen
  \bibfield  {author} {\bibinfo {author} {\bibfnamefont {A.}~\bibnamefont
  {Vladimirov}},\ }\href {\doibase 10.1007/JHEP10(2019)090} {\bibfield
  {journal} {\bibinfo  {journal} {JHEP}\ }\textbf {\bibinfo {volume} {10}},\
  \bibinfo {pages} {090} (\bibinfo {year} {2019})},\ \Eprint
  {http://arxiv.org/abs/1907.10356} {arXiv:1907.10356 [hep-ph]} \BibitemShut
  {NoStop}%
\bibitem [{\citenamefont {Bury}\ \emph {et~al.}(2021)\citenamefont {Bury},
  \citenamefont {Prokudin},\ and\ \citenamefont {Vladimirov}}]{Bury:2020vhj}%
  \BibitemOpen
  \bibfield  {author} {\bibinfo {author} {\bibfnamefont {M.}~\bibnamefont
  {Bury}}, \bibinfo {author} {\bibfnamefont {A.}~\bibnamefont {Prokudin}}, \
  and\ \bibinfo {author} {\bibfnamefont {A.}~\bibnamefont {Vladimirov}},\
  }\href {\doibase 10.1103/PhysRevLett.126.112002} {\bibfield  {journal}
  {\bibinfo  {journal} {Phys. Rev. Lett.}\ }\textbf {\bibinfo {volume} {126}},\
  \bibinfo {pages} {112002} (\bibinfo {year} {2021})},\ \Eprint
  {http://arxiv.org/abs/2012.05135} {arXiv:2012.05135 [hep-ph]} \BibitemShut
  {NoStop}%
\bibitem [{\citenamefont {Moos}\ \emph {et~al.}(2024)\citenamefont {Moos},
  \citenamefont {Scimemi}, \citenamefont {Vladimirov},\ and\ \citenamefont
  {Zurita}}]{Moos:2023yfa}%
  \BibitemOpen
  \bibfield  {author} {\bibinfo {author} {\bibfnamefont {V.}~\bibnamefont
  {Moos}}, \bibinfo {author} {\bibfnamefont {I.}~\bibnamefont {Scimemi}},
  \bibinfo {author} {\bibfnamefont {A.}~\bibnamefont {Vladimirov}}, \ and\
  \bibinfo {author} {\bibfnamefont {P.}~\bibnamefont {Zurita}},\ }\href
  {\doibase 10.1007/JHEP05(2024)036} {\bibfield  {journal} {\bibinfo  {journal}
  {JHEP}\ }\textbf {\bibinfo {volume} {05}},\ \bibinfo {pages} {036} (\bibinfo
  {year} {2024})},\ \Eprint {http://arxiv.org/abs/2305.07473} {arXiv:2305.07473
  [hep-ph]} \BibitemShut {NoStop}%
\bibitem [{\citenamefont {Bacchetta}\ \emph {et~al.}(2022)\citenamefont
  {Bacchetta}, \citenamefont {Bertone}, \citenamefont {Bissolotti},
  \citenamefont {Bozzi}, \citenamefont {Cerutti}, \citenamefont {Piacenza},
  \citenamefont {Radici},\ and\ \citenamefont {Signori}}]{Bacchetta:2022awv}%
  \BibitemOpen
  \bibfield  {author} {\bibinfo {author} {\bibfnamefont {A.}~\bibnamefont
  {Bacchetta}}, \bibinfo {author} {\bibfnamefont {V.}~\bibnamefont {Bertone}},
  \bibinfo {author} {\bibfnamefont {C.}~\bibnamefont {Bissolotti}}, \bibinfo
  {author} {\bibfnamefont {G.}~\bibnamefont {Bozzi}}, \bibinfo {author}
  {\bibfnamefont {M.}~\bibnamefont {Cerutti}}, \bibinfo {author} {\bibfnamefont
  {F.}~\bibnamefont {Piacenza}}, \bibinfo {author} {\bibfnamefont
  {M.}~\bibnamefont {Radici}}, \ and\ \bibinfo {author} {\bibfnamefont
  {A.}~\bibnamefont {Signori}} (\bibinfo {collaboration} {MAP
  (Multi-dimensional Analyses of Partonic distributions)}),\ }\href {\doibase
  10.1007/JHEP10(2022)127} {\bibfield  {journal} {\bibinfo  {journal} {JHEP}\
  }\textbf {\bibinfo {volume} {10}},\ \bibinfo {pages} {127} (\bibinfo {year}
  {2022})},\ \Eprint {http://arxiv.org/abs/2206.07598} {arXiv:2206.07598
  [hep-ph]} \BibitemShut {NoStop}%
\bibitem [{\citenamefont {Boglione}\ \emph {et~al.}(2022)\citenamefont
  {Boglione}, \citenamefont {Gonzalez-Hernandez},\ and\ \citenamefont
  {Simonelli}}]{Boglione:2022nzq}%
  \BibitemOpen
  \bibfield  {author} {\bibinfo {author} {\bibfnamefont {M.}~\bibnamefont
  {Boglione}}, \bibinfo {author} {\bibfnamefont {J.~O.}\ \bibnamefont
  {Gonzalez-Hernandez}}, \ and\ \bibinfo {author} {\bibfnamefont
  {A.}~\bibnamefont {Simonelli}},\ }\href {\doibase
  10.1103/PhysRevD.106.074024} {\bibfield  {journal} {\bibinfo  {journal}
  {Phys. Rev. D}\ }\textbf {\bibinfo {volume} {106}},\ \bibinfo {pages}
  {074024} (\bibinfo {year} {2022})},\ \Eprint
  {http://arxiv.org/abs/2206.08876} {arXiv:2206.08876 [hep-ph]} \BibitemShut
  {NoStop}%
\bibitem [{\citenamefont {Barry}\ \emph {et~al.}(2023)\citenamefont {Barry},
  \citenamefont {Gamberg}, \citenamefont {Melnitchouk}, \citenamefont {Moffat},
  \citenamefont {Pitonyak}, \citenamefont {Prokudin},\ and\ \citenamefont
  {Sato}}]{Barry:2023qqh}%
  \BibitemOpen
  \bibfield  {author} {\bibinfo {author} {\bibfnamefont {P.~C.}\ \bibnamefont
  {Barry}}, \bibinfo {author} {\bibfnamefont {L.}~\bibnamefont {Gamberg}},
  \bibinfo {author} {\bibfnamefont {W.}~\bibnamefont {Melnitchouk}}, \bibinfo
  {author} {\bibfnamefont {E.}~\bibnamefont {Moffat}}, \bibinfo {author}
  {\bibfnamefont {D.}~\bibnamefont {Pitonyak}}, \bibinfo {author}
  {\bibfnamefont {A.}~\bibnamefont {Prokudin}}, \ and\ \bibinfo {author}
  {\bibfnamefont {N.}~\bibnamefont {Sato}} (\bibinfo {collaboration} {Jefferson
  Lab Angular Momentum (JAM)}),\ }\href {\doibase 10.1103/PhysRevD.108.L091504}
  {\bibfield  {journal} {\bibinfo  {journal} {Phys. Rev. D}\ }\textbf {\bibinfo
  {volume} {108}},\ \bibinfo {pages} {L091504} (\bibinfo {year} {2023})},\
  \Eprint {http://arxiv.org/abs/2302.01192} {arXiv:2302.01192 [hep-ph]}
  \BibitemShut {NoStop}%
\bibitem [{\citenamefont {Boglione}\ and\ \citenamefont
  {Simonelli}(2023)}]{Boglione:2023duo}%
  \BibitemOpen
  \bibfield  {author} {\bibinfo {author} {\bibfnamefont {M.}~\bibnamefont
  {Boglione}}\ and\ \bibinfo {author} {\bibfnamefont {A.}~\bibnamefont
  {Simonelli}},\ }\href {\doibase 10.1007/JHEP09(2023)006} {\bibfield
  {journal} {\bibinfo  {journal} {JHEP}\ }\textbf {\bibinfo {volume} {09}},\
  \bibinfo {pages} {006} (\bibinfo {year} {2023})},\ \Eprint
  {http://arxiv.org/abs/2306.02937} {arXiv:2306.02937 [hep-ph]} \BibitemShut
  {NoStop}%
\bibitem [{\citenamefont {Bacchetta}\ \emph {et~al.}(2024)\citenamefont
  {Bacchetta}, \citenamefont {Bertone}, \citenamefont {Bissolotti},
  \citenamefont {Bozzi}, \citenamefont {Cerutti}, \citenamefont {Delcarro},
  \citenamefont {Radici}, \citenamefont {Rossi},\ and\ \citenamefont
  {Signori}}]{Bacchetta:2024qre}%
  \BibitemOpen
  \bibfield  {author} {\bibinfo {author} {\bibfnamefont {A.}~\bibnamefont
  {Bacchetta}}, \bibinfo {author} {\bibfnamefont {V.}~\bibnamefont {Bertone}},
  \bibinfo {author} {\bibfnamefont {C.}~\bibnamefont {Bissolotti}}, \bibinfo
  {author} {\bibfnamefont {G.}~\bibnamefont {Bozzi}}, \bibinfo {author}
  {\bibfnamefont {M.}~\bibnamefont {Cerutti}}, \bibinfo {author} {\bibfnamefont
  {F.}~\bibnamefont {Delcarro}}, \bibinfo {author} {\bibfnamefont
  {M.}~\bibnamefont {Radici}}, \bibinfo {author} {\bibfnamefont
  {L.}~\bibnamefont {Rossi}}, \ and\ \bibinfo {author} {\bibfnamefont
  {A.}~\bibnamefont {Signori}} (\bibinfo {collaboration} {MAP}),\ }\href
  {\doibase 10.1007/JHEP08(2024)232} {\bibfield  {journal} {\bibinfo  {journal}
  {JHEP}\ }\textbf {\bibinfo {volume} {08}},\ \bibinfo {pages} {232} (\bibinfo
  {year} {2024})},\ \Eprint {http://arxiv.org/abs/2405.13833} {arXiv:2405.13833
  [hep-ph]} \BibitemShut {NoStop}%
\bibitem [{\citenamefont {Barry}\ \emph {et~al.}(2025)\citenamefont {Barry}
  \emph {et~al.}}]{Barry:2025glq}%
  \BibitemOpen
  \bibfield  {author} {\bibinfo {author} {\bibfnamefont {P.~C.}\ \bibnamefont
  {Barry}} \emph {et~al.},\ }\href@noop {} {\  (\bibinfo {year} {2025})},\
  \Eprint {http://arxiv.org/abs/2510.13771} {arXiv:2510.13771 [hep-ph]}
  \BibitemShut {NoStop}%
\bibitem [{\citenamefont {Avkhadiev}\ \emph {et~al.}(2025)\citenamefont
  {Avkhadiev}, \citenamefont {Bertone}, \citenamefont {Bissolotti},
  \citenamefont {Cerutti}, \citenamefont {Fu}, \citenamefont {Rodini},
  \citenamefont {Shanahan}, \citenamefont {Wagman},\ and\ \citenamefont
  {Zhao}}]{Avkhadiev:2025wps}%
  \BibitemOpen
  \bibfield  {author} {\bibinfo {author} {\bibfnamefont {A.}~\bibnamefont
  {Avkhadiev}}, \bibinfo {author} {\bibfnamefont {V.}~\bibnamefont {Bertone}},
  \bibinfo {author} {\bibfnamefont {C.}~\bibnamefont {Bissolotti}}, \bibinfo
  {author} {\bibfnamefont {M.}~\bibnamefont {Cerutti}}, \bibinfo {author}
  {\bibfnamefont {Y.}~\bibnamefont {Fu}}, \bibinfo {author} {\bibfnamefont
  {S.}~\bibnamefont {Rodini}}, \bibinfo {author} {\bibfnamefont
  {P.}~\bibnamefont {Shanahan}}, \bibinfo {author} {\bibfnamefont
  {M.}~\bibnamefont {Wagman}}, \ and\ \bibinfo {author} {\bibfnamefont
  {Y.}~\bibnamefont {Zhao}},\ }\href@noop {} {\  (\bibinfo {year} {2025})},\
  \Eprint {http://arxiv.org/abs/2510.26489} {arXiv:2510.26489 [hep-ph]}
  \BibitemShut {NoStop}%
\bibitem [{\citenamefont {Balitsky}(2021)}]{Balitsky:2020jzt}%
  \BibitemOpen
  \bibfield  {author} {\bibinfo {author} {\bibfnamefont {I.}~\bibnamefont
  {Balitsky}},\ }\href {\doibase 10.1007/JHEP05(2021)046} {\bibfield  {journal}
  {\bibinfo  {journal} {JHEP}\ }\textbf {\bibinfo {volume} {05}},\ \bibinfo
  {pages} {046} (\bibinfo {year} {2021})},\ \Eprint
  {http://arxiv.org/abs/2012.01588} {arXiv:2012.01588 [hep-ph]} \BibitemShut
  {NoStop}%
\bibitem [{\citenamefont {Vladimirov}\ \emph {et~al.}(2022)\citenamefont
  {Vladimirov}, \citenamefont {Moos},\ and\ \citenamefont
  {Scimemi}}]{Vladimirov:2021hdn}%
  \BibitemOpen
  \bibfield  {author} {\bibinfo {author} {\bibfnamefont {A.}~\bibnamefont
  {Vladimirov}}, \bibinfo {author} {\bibfnamefont {V.}~\bibnamefont {Moos}}, \
  and\ \bibinfo {author} {\bibfnamefont {I.}~\bibnamefont {Scimemi}},\ }\href
  {\doibase 10.1007/JHEP01(2022)110} {\bibfield  {journal} {\bibinfo  {journal}
  {JHEP}\ }\textbf {\bibinfo {volume} {01}},\ \bibinfo {pages} {110} (\bibinfo
  {year} {2022})},\ \Eprint {http://arxiv.org/abs/2109.09771} {arXiv:2109.09771
  [hep-ph]} \BibitemShut {NoStop}%
\bibitem [{\citenamefont {Vladimirov}(2023)}]{Vladimirov:2023aot}%
  \BibitemOpen
  \bibfield  {author} {\bibinfo {author} {\bibfnamefont {A.}~\bibnamefont
  {Vladimirov}},\ }\href {\doibase 10.1007/JHEP12(2023)008} {\bibfield
  {journal} {\bibinfo  {journal} {JHEP}\ }\textbf {\bibinfo {volume} {12}},\
  \bibinfo {pages} {008} (\bibinfo {year} {2023})},\ \Eprint
  {http://arxiv.org/abs/2307.13054} {arXiv:2307.13054 [hep-ph]} \BibitemShut
  {NoStop}%
\bibitem [{\citenamefont {Vladimirov}\ \emph {et~al.}(2025)\citenamefont
  {Vladimirov}, \citenamefont {Portela},\ and\ \citenamefont
  {Rodini}}]{Vladimirov:2025qrh}%
  \BibitemOpen
  \bibfield  {author} {\bibinfo {author} {\bibfnamefont {A.}~\bibnamefont
  {Vladimirov}}, \bibinfo {author} {\bibfnamefont {G.}~\bibnamefont {Portela}},
  \ and\ \bibinfo {author} {\bibfnamefont {S.}~\bibnamefont {Rodini}},\
  }\href@noop {} {\  (\bibinfo {year} {2025})},\ \Eprint
  {http://arxiv.org/abs/2511.04294} {arXiv:2511.04294 [hep-ph]} \BibitemShut
  {NoStop}%
\bibitem [{\citenamefont {Arroyo-Castro}\ \emph {et~al.}(2025)\citenamefont
  {Arroyo-Castro}, \citenamefont {Scimemi},\ and\ \citenamefont
  {Vladimirov}}]{Arroyo-Castro:2025slx}%
  \BibitemOpen
  \bibfield  {author} {\bibinfo {author} {\bibfnamefont {A.}~\bibnamefont
  {Arroyo-Castro}}, \bibinfo {author} {\bibfnamefont {I.}~\bibnamefont
  {Scimemi}}, \ and\ \bibinfo {author} {\bibfnamefont {A.}~\bibnamefont
  {Vladimirov}},\ }\href {\doibase 10.1007/JHEP06(2025)202} {\bibfield
  {journal} {\bibinfo  {journal} {JHEP}\ }\textbf {\bibinfo {volume} {06}},\
  \bibinfo {pages} {202} (\bibinfo {year} {2025})},\ \Eprint
  {http://arxiv.org/abs/2503.24336} {arXiv:2503.24336 [hep-ph]} \BibitemShut
  {NoStop}%
\bibitem [{\citenamefont {Braun}\ \emph {et~al.}(2019)\citenamefont {Braun},
  \citenamefont {Vladimirov},\ and\ \citenamefont {Zhang}}]{Braun:2018brg}%
  \BibitemOpen
  \bibfield  {author} {\bibinfo {author} {\bibfnamefont {V.~M.}\ \bibnamefont
  {Braun}}, \bibinfo {author} {\bibfnamefont {A.}~\bibnamefont {Vladimirov}}, \
  and\ \bibinfo {author} {\bibfnamefont {J.-H.}\ \bibnamefont {Zhang}},\ }\href
  {\doibase 10.1103/PhysRevD.99.014013} {\bibfield  {journal} {\bibinfo
  {journal} {Phys. Rev. D}\ }\textbf {\bibinfo {volume} {99}},\ \bibinfo
  {pages} {014013} (\bibinfo {year} {2019})},\ \Eprint
  {http://arxiv.org/abs/1810.00048} {arXiv:1810.00048 [hep-ph]} \BibitemShut
  {NoStop}%
\bibitem [{\citenamefont {Rodini}\ and\ \citenamefont
  {Vladimirov}(2022)}]{Rodini:2022wki}%
  \BibitemOpen
  \bibfield  {author} {\bibinfo {author} {\bibfnamefont {S.}~\bibnamefont
  {Rodini}}\ and\ \bibinfo {author} {\bibfnamefont {A.}~\bibnamefont
  {Vladimirov}},\ }\href@noop {} {\  (\bibinfo {year} {2022})},\ \Eprint
  {http://arxiv.org/abs/2204.03856} {arXiv:2204.03856 [hep-ph]} \BibitemShut
  {NoStop}%
\bibitem [{\citenamefont {Manohar}\ \emph {et~al.}(2002)\citenamefont
  {Manohar}, \citenamefont {Mehen}, \citenamefont {Pirjol},\ and\ \citenamefont
  {Stewart}}]{Manohar:2002fd}%
  \BibitemOpen
  \bibfield  {author} {\bibinfo {author} {\bibfnamefont {A.~V.}\ \bibnamefont
  {Manohar}}, \bibinfo {author} {\bibfnamefont {T.}~\bibnamefont {Mehen}},
  \bibinfo {author} {\bibfnamefont {D.}~\bibnamefont {Pirjol}}, \ and\ \bibinfo
  {author} {\bibfnamefont {I.~W.}\ \bibnamefont {Stewart}},\ }\href {\doibase
  10.1016/S0370-2693(02)02029-4} {\bibfield  {journal} {\bibinfo  {journal}
  {Phys. Lett. B}\ }\textbf {\bibinfo {volume} {539}},\ \bibinfo {pages} {59}
  (\bibinfo {year} {2002})},\ \Eprint {http://arxiv.org/abs/hep-ph/0204229}
  {arXiv:hep-ph/0204229} \BibitemShut {NoStop}%
\bibitem [{\citenamefont {Piloneta}\ and\ \citenamefont
  {Vladimirov}(2024)}]{Piloneta:2024aac}%
  \BibitemOpen
  \bibfield  {author} {\bibinfo {author} {\bibfnamefont {S.}~\bibnamefont
  {Piloneta}}\ and\ \bibinfo {author} {\bibfnamefont {A.}~\bibnamefont
  {Vladimirov}},\ }\href {\doibase 10.1007/JHEP12(2024)059} {\bibfield
  {journal} {\bibinfo  {journal} {JHEP}\ }\textbf {\bibinfo {volume} {12}},\
  \bibinfo {pages} {059} (\bibinfo {year} {2024})},\ \Eprint
  {http://arxiv.org/abs/2407.06277} {arXiv:2407.06277 [hep-ph]} \BibitemShut
  {NoStop}%
\bibitem [{\citenamefont {Braun}\ and\ \citenamefont
  {Manashov}(2012)}]{Braun:2011dg}%
  \BibitemOpen
  \bibfield  {author} {\bibinfo {author} {\bibfnamefont {V.~M.}\ \bibnamefont
  {Braun}}\ and\ \bibinfo {author} {\bibfnamefont {A.~N.}\ \bibnamefont
  {Manashov}},\ }\href {\doibase 10.1007/JHEP01(2012)085} {\bibfield  {journal}
  {\bibinfo  {journal} {JHEP}\ }\textbf {\bibinfo {volume} {01}},\ \bibinfo
  {pages} {085} (\bibinfo {year} {2012})},\ \Eprint
  {http://arxiv.org/abs/1111.6765} {arXiv:1111.6765 [hep-ph]} \BibitemShut
  {NoStop}%
\bibitem [{\citenamefont {Braun}\ and\ \citenamefont
  {Manashov}(2011)}]{Braun:2011zr}%
  \BibitemOpen
  \bibfield  {author} {\bibinfo {author} {\bibfnamefont {V.~M.}\ \bibnamefont
  {Braun}}\ and\ \bibinfo {author} {\bibfnamefont {A.~N.}\ \bibnamefont
  {Manashov}},\ }\href {\doibase 10.1103/PhysRevLett.107.202001} {\bibfield
  {journal} {\bibinfo  {journal} {Phys. Rev. Lett.}\ }\textbf {\bibinfo
  {volume} {107}},\ \bibinfo {pages} {202001} (\bibinfo {year} {2011})},\
  \Eprint {http://arxiv.org/abs/1108.2394} {arXiv:1108.2394 [hep-ph]}
  \BibitemShut {NoStop}%
\bibitem [{\citenamefont {Piloneta}\ and\ \citenamefont
  {Vladimirov}(2025)}]{Piloneta:2025jjb}%
  \BibitemOpen
  \bibfield  {author} {\bibinfo {author} {\bibfnamefont {S.}~\bibnamefont
  {Piloneta}}\ and\ \bibinfo {author} {\bibfnamefont {A.}~\bibnamefont
  {Vladimirov}},\ }\href@noop {} {\  (\bibinfo {year} {2025})},\ \Eprint
  {http://arxiv.org/abs/2510.14496} {arXiv:2510.14496 [hep-ph]} \BibitemShut
  {NoStop}%
\bibitem [{\citenamefont {del Rio}\ \emph {et~al.}(2024)\citenamefont {del
  Rio}, \citenamefont {Prokudin}, \citenamefont {Scimemi},\ and\ \citenamefont
  {Vladimirov}}]{delRio:2024vvq}%
  \BibitemOpen
  \bibfield  {author} {\bibinfo {author} {\bibfnamefont {O.}~\bibnamefont {del
  Rio}}, \bibinfo {author} {\bibfnamefont {A.}~\bibnamefont {Prokudin}},
  \bibinfo {author} {\bibfnamefont {I.}~\bibnamefont {Scimemi}}, \ and\
  \bibinfo {author} {\bibfnamefont {A.}~\bibnamefont {Vladimirov}},\ }\href
  {\doibase 10.1103/PhysRevD.110.016003} {\bibfield  {journal} {\bibinfo
  {journal} {Phys. Rev. D}\ }\textbf {\bibinfo {volume} {110}},\ \bibinfo
  {pages} {016003} (\bibinfo {year} {2024})},\ \Eprint
  {http://arxiv.org/abs/2402.01836} {arXiv:2402.01836 [hep-ph]} \BibitemShut
  {NoStop}%
\bibitem [{\citenamefont {del R{\'\i}o}\ and\ \citenamefont
  {Vladimirov}(2023)}]{delRio:2023pse}%
  \BibitemOpen
  \bibfield  {author} {\bibinfo {author} {\bibfnamefont {{\'O}.}~\bibnamefont
  {del R{\'\i}o}}\ and\ \bibinfo {author} {\bibfnamefont {A.}~\bibnamefont
  {Vladimirov}},\ }\href {\doibase 10.1103/PhysRevD.108.114009} {\bibfield
  {journal} {\bibinfo  {journal} {Phys. Rev. D}\ }\textbf {\bibinfo {volume}
  {108}},\ \bibinfo {pages} {114009} (\bibinfo {year} {2023})},\ \Eprint
  {http://arxiv.org/abs/2304.14440} {arXiv:2304.14440 [hep-ph]} \BibitemShut
  {NoStop}%
\bibitem [{\citenamefont {Ji}\ \emph {et~al.}(2023)\citenamefont {Ji},
  \citenamefont {Liu},\ and\ \citenamefont {Su}}]{Ji:2023pba}%
  \BibitemOpen
  \bibfield  {author} {\bibinfo {author} {\bibfnamefont {X.}~\bibnamefont
  {Ji}}, \bibinfo {author} {\bibfnamefont {Y.}~\bibnamefont {Liu}}, \ and\
  \bibinfo {author} {\bibfnamefont {Y.}~\bibnamefont {Su}},\ }\href {\doibase
  10.1007/JHEP08(2023)037} {\bibfield  {journal} {\bibinfo  {journal} {JHEP}\
  }\textbf {\bibinfo {volume} {08}},\ \bibinfo {pages} {037} (\bibinfo {year}
  {2023})},\ \Eprint {http://arxiv.org/abs/2305.04416} {arXiv:2305.04416
  [hep-ph]} \BibitemShut {NoStop}%
\bibitem [{\citenamefont {Liu}\ and\ \citenamefont {Su}(2024)}]{Liu:2023onm}%
  \BibitemOpen
  \bibfield  {author} {\bibinfo {author} {\bibfnamefont {Y.}~\bibnamefont
  {Liu}}\ and\ \bibinfo {author} {\bibfnamefont {Y.}~\bibnamefont {Su}},\
  }\href {\doibase 10.1007/JHEP02(2024)204} {\bibfield  {journal} {\bibinfo
  {journal} {JHEP}\ }\textbf {\bibinfo {volume} {2024}},\ \bibinfo {pages}
  {204} (\bibinfo {year} {2024})},\ \Eprint {http://arxiv.org/abs/2311.06907}
  {arXiv:2311.06907 [hep-ph]} \BibitemShut {NoStop}%
\bibitem [{\citenamefont {Moos}\ \emph {et~al.}(2025)\citenamefont {Moos},
  \citenamefont {Scimemi}, \citenamefont {Vladimirov},\ and\ \citenamefont
  {Zurita}}]{Moos:2025sal}%
  \BibitemOpen
  \bibfield  {author} {\bibinfo {author} {\bibfnamefont {V.}~\bibnamefont
  {Moos}}, \bibinfo {author} {\bibfnamefont {I.}~\bibnamefont {Scimemi}},
  \bibinfo {author} {\bibfnamefont {A.}~\bibnamefont {Vladimirov}}, \ and\
  \bibinfo {author} {\bibfnamefont {P.}~\bibnamefont {Zurita}},\ }\href
  {\doibase 10.1007/JHEP11(2025)134} {\bibfield  {journal} {\bibinfo  {journal}
  {JHEP}\ }\textbf {\bibinfo {volume} {11}},\ \bibinfo {pages} {134} (\bibinfo
  {year} {2025})},\ \Eprint {http://arxiv.org/abs/2503.11201} {arXiv:2503.11201
  [hep-ph]} \BibitemShut {NoStop}%
\end{thebibliography}%

\end{document}